\documentclass[11pt]{article}

\usepackage{physics}
\usepackage{braket}
\usepackage{mathtools}
\usepackage[margin=1in]{geometry}
\usepackage{amsmath,amssymb,amsfonts,setspace,url,mathrsfs}
\usepackage{booktabs}
\usepackage{bm}
\usepackage{algpseudocode}
\usepackage{enumitem}
\usepackage{latexsym}
\usepackage{amsthm}
\usepackage{subfigure}
\usepackage{float}
\usepackage{stmaryrd}
\usepackage{fancybox}
\usepackage{bigstrut,array,multirow}
\usepackage{here}
\usepackage{color}
\usepackage{url}
\usepackage{hyperref}
\usepackage[capitalise]{cleveref}
\usepackage{mathpazo}
\usepackage{thmtools}
\usepackage{thm-restate}
\usepackage[normalem]{ulem}

\usepackage{xcolor}
\usepackage{tikz}
\usetikzlibrary{arrows.meta}

\newtheorem{theorem}{Theorem}
\newtheorem{corollary}[theorem]{Corollary}

\newtheorem{claim}[theorem]{Claim}
\newtheorem{observation}[theorem]{Observation}
\newtheorem{lemma}[theorem]{Lemma}
\newtheorem{definition}[theorem]{Definition}

\newcommand{\dist}{\text{dist}}

\newcommand{\diam}{{\text{diam}}}

\newcommand{\Pess}{D}

\newcommand{\LL}{\Lambda_H^*}
\newcommand{\GG}{\Lambda_H'}
\newcommand{\RR}{R}
\newcommand{\CC}{\mathcal C}

\newcounter{note}

\newcommand{\myparskip}{3pt}
\parskip \myparskip
\newcommand{\Oh}{O}

\crefname{enumi}{step}{step}
\Crefname{enumi}{Step}{Step}

\begin{document}

	\title{Distance Preserver and Bridge Girth}

\begin{titlepage}
	
	\title{Improved Bounds on Distance Preservers}

\author{Yanyu Chen\thanks{National University of Singapore, Singapore. Email: {\tt yanyu.chen@u.nus.edu}.}
\and Yu Chen\thanks{National University of Singapore, Singapore. Email: {\tt yu.chen@nus.edu.sg}.} 
\and Zihan Tan\thanks{University of Minnesota Twin Cities, MN, USA. Email: {\tt ztan@umn.edu}.} \and Mingyang Yang\thanks{National University of Singapore, Singapore. Email: {\tt myangat@u.nus.edu}.}} 

	\maketitle
	
	\thispagestyle{empty}
	\begin{abstract}
Given an undirected unweighted graph \(G\) on $n$ vertices and a set \(P\) of $p$ pairs of its vertices, a distance preserver is a sparse subgraph of $G$ that exactly preserves the shortest-path distances between all pairs in \(P\).  In this paper, we prove a new upper bound of $O\big(n^{8/9} p^{4/9}\big)$ for distance preservers when $p\in [n^{1/2}, n]$. Our result improves the previous bound of $O\big(n^{2/3}p^{2/3}+np^{1/3}\big)$ by Bodwin and  Vassilevska Williams and $O\big(\min\{n+n^{1/2}p,\,np^{1/2}\}\big)$ by Coppersmith and Elkin, particularly in the regime $p\in [n^{7/10}, n]$, and narrows the gap toward the conjectured \(O(n^{2/3}p^{2/3}+n)\) bound.

Using the techniques developed for distance preservers, we are also able to obtain pairwise additive $+2$ spanners with $O(n+n^{9/10}p^{2/5})$ edges, pairwise additive $+4$ spanners with $O(n+n^{24/25}p^{8/25})$ edges, and new bounds for sourcewise spanners, improving upon previous results in certain parameter regimes.

    \end{abstract}
\end{titlepage}

\tableofcontents

\newpage

\section{Introduction}

Let $G=(V,E)$ be an $n$-node undirected unweighted graph and let $P\subseteq V\times V$ be a set of $p=|P|$ \emph{demand pairs}. A subgraph $H=(V,E_H)$, where $E_H\subseteq E$, is called a \emph{distance preserver} for $(G,P)$ if
\[
    \dist_H(s,t)=\dist_G(s,t) \qquad \text{for every } (s,t)\in P.
\]
The goal is to determine its worst-case existential bound as a function of $n$ and $p$, namely the minimum worst-case number of edges that always suffices for a distance preserver.


As a central object in graph sparsification/compression introduced by Coppersmith and Elkin~\cite{CE06}, distance preservers have been studied extensively.
By definition, for a graph $G$ and a set $P$ of demand pairs, a distance preserver for $(G,P)$ is essentially the union of, for each pair in $P$, a shortest path in $G$ connecting them. The focus is how to choose these shortest paths, which we call a \emph{tie-breaking scheme}.
Coppersmith and Elkin~\cite{CE06} showed that a basic property a tiebreaking scheme can have is \emph{consistency}, meaning that the intersection of any
two chosen shortest paths is either empty or a subpath of both.
With this property, they derived the size bound
$O\!\left(\min\{n+n^{1/2}p,\,np^{1/2}\}\right)$. 
When $p=O(n^{1/2})$, their bound is $O(n)$ and is clearly optimal. The remaining interesting regime is when $p\in [n^{1/2},n^2]$.

A major advance was obtained by Bodwin and Vassilevska Williams~\cite{BVW21}. They showed that consistency alone is not sufficient to obtain the optimal bound: there are consistent tie-breaking schemes whose unions achieve the Coppersmith-Elkin bound, but a different choice of shortest paths can be polynomially better. With further investigation on how the chosen shortest paths may branch with each other, they improved the upper bound to
$O\!\left(n^{2/3}p^{2/3}+np^{1/3}\right)$, which remained the best upper bound to date.
Furthermore, they conjectured that it could be $O\!\left(n^{2/3}p^{2/3}+n\right)$. In other words, in the regime $p\in [n^{1/2},n]$, the bound is conjectured to be $O(n^{2/3}p^{2/3})$, instead of only $O(np^{1/3})$.

In this work, we make progress towards proving their conjecture. Our result can be summarized as the following theorem.

\begin{theorem}
\label{thm: main}

Let $G$ be any $n$-node undirected unweighted graph and let $P$ be any set of $p$ demand pairs with $n^{1/2}\le p\le n$. Then $(G,P)$ admits a distance preserver of $O\!\left(n^{8/9} p^{4/9}\right)$ edges.
\end{theorem}

\begin{figure}[h]
\centering
\includegraphics[width=0.72\linewidth]{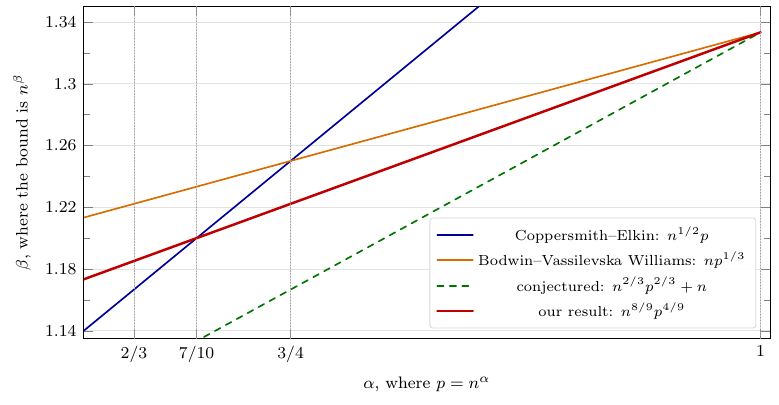}
\caption{Comparison between our results and previous work.\label{fig: comparison}}
\end{figure}

Compared with previous results in \cite{CE06} and \cite{BVW21}, our bound achieves an improvement in the regime $p\in [n^{7/10}, n]$. See \Cref{fig: comparison} for comparison.

\paragraph{Applications to additive spanners.}
Our techniques also yield new bounds for pairwise and sourcewise additive
spanners. A pairwise additive $+\beta$ spanner for $(G,P)$ is a subgraph
$H\subseteq G$ satisfying
$\operatorname{dist}_H(x,y)\leq \operatorname{dist}_G(x,y)+\beta$
for every $(x,y)\in P$.
For arbitrary demand pairs, we obtain a $+2$ spanner with
$O(n+n^{9/10}p^{2/5})$ edges for all $p$, and a $+4$ spanner with
$O(n+n^{24/25}p^{8/25})$ edges when $p\leq n^{7/6}$.
We further obtain bounds that exploit a restricted source set:
when $P\subseteq S\times V$ and $\sigma=|S|$, a common sampling reduction
from our exact-preserver bound gives $+2$, $+4$, and $+6$ spanners whose
sizes depend jointly on $n$, $p$, and $\sigma$.
When all reachable pairs in $S\times V$ are demanded, an additional
clustering reduction gives stronger complete-sourcewise guarantees.
Table~\ref{tab:additive-spanners} summarizes these bounds, together with
a further sourcewise $+6$ consequence of the classical exact-preserver bound.

\begin{table}[p]
  \centering
  \footnotesize
  \setlength{\tabcolsep}{3pt}
  \setlength{\jot}{0pt}
  \renewcommand{\arraystretch}{1.05}
  \newcommand{\boundmin}[1]{%
    $\displaystyle\min\!\left\{\begin{gathered}#1\end{gathered}\right\}$}
  \newcommand{\prevbound}[2]{#1\;\text{\cite{#2}}}
  \resizebox{\textwidth}{!}{%
  \begin{tabular}{@{}clllll@{}}
    \toprule
    Error & Previous bounds & Our bound & Range
      & Improvement range & Result \\
    \midrule
    \multicolumn{6}{@{}l}{\textbf{Pairwise: arbitrary $P\subseteq V\times V$}} \\
    \addlinespace[3pt]
    $+2$
      & \boundmin{\prevbound{n^{1/2}p}{CE06},\quad\prevbound{np^{1/3}}{Bod22},\\
                   \prevbound{n^{3/2}}{EP04}}
      & $n^{9/10}p^{2/5}$ & All $p$
      & $n^{2/3}<p<n^{3/2}$
      & \Cref{thm: main +2 spanner} \\
    \addlinespace[3pt]
    $+4$
      & \boundmin{\prevbound{n^{1/2}p}{CE06},\quad\prevbound{np^{2/7}}{Bod22},\\
                   \prevbound{n^{9/10}p^{2/5}}{CE06},\\
                   \prevbound{n^{7/5}}{Che13}}
      & $n^{24/25}p^{8/25}$ & $p\leq n^{7/6}$
      & $n^{3/4}<p<n^{7/6}$
      & \Cref{thm: main +4 spanner} \\
    \midrule
    \multicolumn{6}{@{}l}{\textbf{Partially specified sourcewise: $P\subseteq S\times V$}} \\
    \addlinespace[3pt]
    $+2$
      & \boundmin{\prevbound{n^{1/2}p}{CE06},\quad\prevbound{np^{1/3}}{Bod22},\\
                   n\sigma\;\text{(BFS trees)},\\
                   \prevbound{n^{5/6}(p\sigma)^{1/3}}{CE06},\\
                   \prevbound{n(p\sigma)^{1/5}}{BVW21}}
      & $n^{16/17}(p\sigma)^{4/17}$
      & $p\sigma\leq n^{5/3}$
      & $\begin{gathered}
          \frac{11}{10}<a+b<\frac53,\\
          26a-8b>15,\\
          12b<3+5a,\\
          13b>4a-1
        \end{gathered}$
      & \Cref{thm: main +2 spanner source-sensitive} \\
    \addlinespace[3pt]
    $+4$
      & \boundmin{\prevbound{n^{1/2}p}{CE06},\quad\prevbound{np^{2/7}}{Bod22},\\
                   n\sigma\;\text{(BFS trees)},\\
                   \prevbound{n^{9/10}p^{2/5}}{CE06},\\
                   \prevbound{n^{9/10}(p\sigma^2)^{1/5}}{CE06},\\
                   \prevbound{n(p\sigma^2)^{1/7}}{BVW21},\\
                   \prevbound{n^{7/6}\sigma^{1/3}}{Par14},\\
                   \prevbound{n\max\{\sigma^{1/2},p^{1/4}\}}{Bod22}}
      & $n^{24/25}(p\sigma^2)^{4/25}$
      & $p\sigma^2\leq n^{7/3}$
      & $\begin{gathered}
          a+2b>\frac32,\\
          17b>4a-1,\\
          32b<4+9a,\\
          24a-2b<31
        \end{gathered}$
      & \Cref{thm: source-sensitive additive} \\
    \addlinespace[3pt]
    $+6$
      & \boundmin{\prevbound{n^{1/2}p}{CE06},\quad\prevbound{np^{1/4}}{Bod22},\\
                   n\sigma\;\text{(BFS trees)},\\
                   \prevbound{n^{13/14}(p\sigma^3)^{1/7}}{CE06},\\
                   \prevbound{n(p\sigma^3)^{1/9}}{BVW21},\\
                   \prevbound{n^{9/8}\sigma^{3/8}}{Par14},\\
                   \prevbound{n^{6/5}\sigma^{1/5}}{Kav17},\\
                   \prevbound{n(\sigma p)^{1/6}}{Kav17}}
      & $n^{32/33}(p\sigma^3)^{4/33}$
      & $p\sigma^3\leq n^3$
      & $\begin{gathered}
          a+3b>\frac{19}{10},\\
          13b<2+3a,\\
          32a-3b<41,\\
          20a+27b<38
        \end{gathered}$
      & \Cref{thm: source-sensitive additive} \\
    \midrule
    \multicolumn{6}{@{}l}{\textbf{Complete sourcewise: $P=S\times V$}} \\
    \addlinespace[3pt]
    $+4$
      & \boundmin{n\sigma\;\text{(BFS trees)},\\
                   \prevbound{n^{7/6}\sigma^{1/3}}{Par14}}
      & $n^{8/7}\sigma^{8/21}$ & $\sigma\leq n^{1/2}$
      & $n^{3/13}<\sigma<n^{1/2}$
      & \Cref{thm: complete-sourcewise additive} \\
    \addlinespace[3pt]
    $+6$
      & \boundmin{n\sigma\;\text{(BFS trees)},\\
                   \prevbound{n^{4/3}}{BKMP10},\\
                   \prevbound{n^{15/14}\sigma^{4/7}}{CE06},\\
                   \prevbound{n^{9/8}\sigma^{3/8}}{Par14},\\
                   \prevbound{n^{6/5}\sigma^{1/5}}{Kav17}}
      & $n^{13/12}\sigma^{1/2}$ & All $\sigma$
      & $n^{1/6}<\sigma<n^{1/3}$
      & \Cref{cor: CE-sourcewise additive} \\
    \addlinespace[3pt]
    $+6$
      & \boundmin{n\sigma\;\text{(BFS trees)},\\
                   \prevbound{n^{15/14}\sigma^{4/7}}{CE06},\\
                   \prevbound{n^{9/8}\sigma^{3/8}}{Par14},\\
                   \prevbound{n^{6/5}\sigma^{1/5}}{Kav17}}
      & $n^{32/29}\sigma^{12/29}$ & $\sigma\leq n^{5/9}$
      & $n^{13/64}<\sigma<n^{14/31}$
      & \Cref{thm: complete-sourcewise additive} \\
    \bottomrule
  \end{tabular}%
  }
  \caption{Additive spanner bounds, where $n=|V|$, $p=|P|$, and
  $\sigma=|S|$. Each entry $f$ denotes $O(n+f)$ edges, with
  polylogarithmic factors suppressed in the previous bounds.
  Previous bounds are simplified on the range in the same row.
  For partial demands, write $p=n^a$ and $\sigma=n^b$, with
  $0\leq b\leq1$ and $b\leq a\leq1+b$; every inequality in a cell
  must hold. Improvement means strict polynomial improvement over the
  previous bounds for fixed exponents, and does not compare different
  new rows against one another.}
  \label{tab:additive-spanners}
\end{table}

\paragraph{Comparison with previous bounds.}
We compare with the exact-preserver bound of~\cite{CE06},
Bodwin's logarithm-free $+2$, $+4$, and $+6$ bounds~\cite{Bod22}, and
classical all-pairs spanners~\cite{EP04,Che13,BKMP10}; the $+4$
all-pairs bound is $\widetilde O(n^{7/5})$~\cite{Che13}.
The sourcewise comparisons additionally use~\cite{KV15,Par14,Kav17}.
We include the consequences of applying the same sampling reductions
 to the classical exact-preserver bounds of~\cite{CE06,BVW21}, as well
as consequences of the subsetwise $+2$ and rectangular $+4$ bounds
in~\cite{Bod22,Kav17}. In particular, $n^{9/10}p^{2/5}$ for pairwise
$+4$ and $n^{15/14}\sigma^{4/7}$ for complete-sourcewise $+6$ are
classical-input consequences, not bounds stated explicitly
in~\cite{CE06}.
All comparisons include an additive $O(n)$ term and take the minimum
of the applicable constructions.

%
%

%
%

\paragraph{Related Work.}
Distance preservers have also been studied in many related forms. Bollobas, Coppersmith, and Elkin studied sparse $D$-preservers and their connection to additive spanners~\cite{BCE05}. Bodwin studied the linear-size regime, directed and weighted variants, large-$p$ behavior, and lower bounds for subset preservers~\cite{Bod21}; related shortest-path structure, including unique shortest-path systems, was further investigated in~\cite{Bod19}. Pairwise spanners and approximate preservers, where one allows multiplicative or additive error on the pairs in $P$, have been studied in~\cite{CGK13,KV15,Kav17,AB16,AB17,ABP17,Neiman24}. Special graph families and graph-metric compression have motivated distance-preserving subgraphs and emulators for planar graphs, interval graphs, and terminal metrics~\cite{AGMW18,CGMW18,GR17,EFN17}. Fault-tolerant distance preservers, which must preserve distances after edge or vertex failures, have been developed in~\cite{PP13,Par15,BGPVW17,GK17,BCPS20,BP23,Parter20,PP25}. Distance preservers also appear in or near the toolkit of distance oracles, additive spanners, hopsets, shortcut sets, and metric compression~\cite{Pet09,BW15,EP16,KP22,KP25,HXX25}. Finally, optimization and algorithmic versions of finding minimum-size or minimum-weight preservers have begun to receive attention~\cite{AYM20,SSV26}.

\paragraph{Organization.} The remainder of the paper is organized as follows. In \Cref{sec: overview}, we start by providing an overview of our techniques. In \Cref{sec: prelim}, we present some formal definitions and important tools from previous work. Then in \Cref{sec: edge-disjoint}, we focus on a special case where the paths chosen by the tie-breaking scheme are edge-disjoint and show an upper bound of $O(n^{5/6}p^{1/2})$. In \Cref{sec:shared}, we establish the technical lemmas needed to handle shared edges in the general case. In \Cref{sec:proof-main}, we combine these lemmas with an iterative demand-removal procedure to prove our main distance-preserver theorem. In \Cref{sec:plus2-pairwise}, we adapt our framework to obtain the pairwise additive $+2$ spanner bound.
Finally, \Cref{sec: +4} presents further results on pairwise
and sourcewise additive spanners.

\section{Technical Overview}
\label{sec: overview}

In this section, we provide a high-level overview of our techniques for proving \Cref{thm: main}.

\subsection*{The Bodwin-Williams framework}

We start by reviewing the approach of proving the $O(n^{2/3}p^{2/3}+np^{1/3})$ bound in \cite{BVW21}. The central notion in their analysis is \emph{branching events}. Given a tie-breaking scheme $\pi$, a branching event occurs when two selected paths enter the same vertex through two different incoming edges.
Coppersmith and Elkin \cite{CE06} showed that if $\pi$ has $b$ branching events, then its distance preserver contains $O(n+\sqrt{nb})$ edges. Therefore, the goal is to prove better bounds for $b$.

Bodwin and Williams showed that beyond \emph{consistency}, a basic property of tie-breaking schemes requiring that any pair of selected paths share a (possibly empty) common subpath, we can also utilize \emph{edge essentiality}, where each edge is essential to some demand pair in the sense that its removal will strictly increase the distance between this pair (note that if an edge is not essential to any pair, we can remove such an edge, only decreasing the size of the preserver). We say that an edge \emph{supports} a demand pair if it is essential to it.
Bodwin and Williams showed that, if many ($\tilde p$) demand pairs are supported by a cluster of ($\sigma$) vertices that are close (at distance at most $\Delta$) to each other, then these demand pairs can be handled together economically, using $\Oh\!\left(n+\sqrt{n\tilde p\sigma\Delta}\right)$ edges.

This bound is then used iteratively. If the current  preserver has average degree at least \(d\), then one can find a vertex of $\Omega(d)$ degree-$\Omega(d)$ neighbors, namely a constant-diameter cluster of $\Theta(d)$ vertices that supports \(\Omega(d^2)\) demand pairs. 
Handling these pairs requires $\Oh(n+\sqrt{nd^3})$ edges.
Remove those demands and repeat the process until the remaining graph has average degree at most \(d\).

In the range \(p\le n\), the main bottleneck is the \(\Oh(n)\) term.  As each round handles roughly \(d^2\) demands, this accumulates to $\frac{p}{d^2}\cdot n$.
Balancing this with the \(\Oh(nd)\) leftover edges in an average-degree-\(d\) graph gives \(d\approx p^{1/3}\) and hence the \(\Oh(np^{1/3})\) term in their bound.

Their framework is nice and clean, but also leaves some room for improvement: if we can manage to handle more than \(d^2\) demands in one neighborhood, we will be able to improve the \(\Oh(np^{1/3})\) bound.

\subsection*{Warm-up: the edge-disjoint case}

We now describe our approach in a special case with the following simplifying assumptions.
\begin{itemize}
\item there is a unique shortest path for each demand pair, and these paths are edge-disjoint; and
\item the degrees of all vertices are roughly the same, around $d$, the average degree.
\end{itemize}
Clearly, in this case each edge $e$ is essential to the only demand pair whose shortest path contains $e$.

We follow the same framework of Bodwin-Williams: find many demand pairs supported by a group of close vertices and handle them together. The main difference is: instead of only looking at constant-diameter neighborhoods of vertices, we explore neighborhoods with larger diameter, and the goal is to handle more than $d^2$ demand pairs in a single round.

On the surface, it is unclear that increasing the diameter will help at all. Consider the BFS tree from any vertex: it seems possible that each level contains $d$ vertices (so the growth is linear) and between consecutive levels there is a bi-clique between them, and all these edges come from the shortest paths of merely $d^2$ demand pairs passing through this neighborhood.

However, we show that this cannot happen.
Let $r$ be a radius parameter.
Consider any vertex $u$. Since (we have assumed that) it has degree $d$ and all its neighbors have degree $d$, the set $P^*$ of demand pairs supported by its neighbors has size at least $|P^*|=\Omega(d^2)$. 
Consider now the radius-$r$ ball of $u$: the shortest path of each of these demand pairs contributes a length-\(\Theta(r)\) subpath inside the ball.  Since the paths are edge-disjoint, this gives \(\Omega(d^2r)\) local edges in the ball.
On the other hand, since these $d^2$ demands are supported by the $d$ neighbors of $u$, applying the BW bound to this neighborhood, we get that the number of local edges is at most \(O(|B(u,r)| +\sqrt{|B(u,r)|\cdot d^3})\).  Comparing with the lower bound \(\Omega(d^2r)\) (and assuming \(r\le d\)), we get that
\[    |B(u,r)|=\Omega(dr^2).\]
Therefore, the neighborhood growth of any vertex cannot be linear, but at least quadratic.

As a consequence, the $r$-neighborhood contains $\Omega(dr^2\cdot d)$ edges. Since each demand pair contributes $O(r)$ edges in this neighborhood (between the first and last time a shortest path touches the ball, the path segment has length at most \(\Oh(r)\)), the $r$-neighborhood supports at least $\Omega(dr^2\cdot d/r)=\Omega(d^2r)$ demands, which is more than $d^2$! 

Applying the BW bound to handle all these demand pairs while balancing the other terms gives an advantage over the $np^{1/3}$ bound. Moreover, it seems that the larger $r$ we take, the greater advantage we will gain. However, there is one place in the discussion above that is not sufficiently rigorous: the shortest paths for the demands in $P^*$ may contribute fewer than $\Omega(d^2 r)$ edges, as some of them may be simply shorter than $r$ themselves. Luckily, these ``short demands'' can be handled separately and cheaply, with at most $r$ edges per demand. We will consequently incur an additional $O(pr)$ term in the distance preserver size bound. Re-balancing all of them, we can prove a bound of $O(n^{5/6}p^{1/2})$, ``half-way'' between the Bodwin-Williams $O(np^{1/3})$ bound and their conjectured $O(n^{2/3}p^{2/3})$ bound.

We present the details for this warm-up case in \Cref{sec: edge-disjoint}.
Remark: the second simplifying assumption (all vertices have degree around $d$) is only for the purpose of technical overview; in \Cref{sec: edge-disjoint} we can get rid of this assumption with some standard techniques.

\subsection*{Handling edge sharing: the general case}

The edge-disjoint assumption is quite powerful since it implies the quadratic growth of any neighborhood, which is also the first thing we lose when the assumption is removed: the set $P^*$ of demand pairs may share edges in their shortest paths, and so the local $r$-neighborhood is no longer guaranteed to contain $\Omega(d^2 r)$ edges, and the remaining analysis falls apart with it.

Therefore, in order to revive the analysis in the edge-disjoint case, it is instructive to first segregate the ``edge-disjoint portion'' of the shortest paths. This brings us to the first central notion: loyal edges.

\subsubsection*{Step 1: Loyal edges, and the assumption that most edges are loyal}
Recall that every edge is essential: removing the edge will strictly increase the distance between some demand pair. If an edge is essential to only one demand pair, then this edge is its ``private property'', and therefore the ``edge-disjoint portion''. In fact, we can be more aggressive here: we say that an edge $(u,v)$ is \emph{loyal} if it is private and both $u$ and $v$ are incident to at least $\Omega(d)$ other private edges (in this case we say both $u$ and $v$ are \emph{rich}), where $d$ is the average degree of the graph.

Although we cannot assume that all edges are loyal, it turns out that we only need to consider the case where most edges are loyal. This is because, intuitively
\begin{itemize}
    \item if a demand pair possesses few private edges, then it can be ignored (after the other demands are handled, we can return to handle them cheaply by adding only their private edges); therefore we can assume that most demand pairs possess many private edges, and so most edges are private; and
    \item we then perform the following pruning: start with the subgraph induced by private edges (such a graph has average degree $\bar d=\Omega(d)$), and repeatedly remove vertices with degree $\le \bar d/10$. Eventually, the total number of removed edges is below $n\bar d/10$, a small fraction of the private edges, and so in the remaining graph, all vertices are rich and all edges are loyal.
\end{itemize}

For the purpose of technical overview, we will make yet another simplifying assumption: \emph{every vertex is incident to $d$ loyal edges and is therefore rich}.
With this simplification, the remaining analysis will be centered around loyal edges.
For a vertex \(u\) and a radius parameter $r$, we define $D(u,r)$ as the set of all demand pairs that possess a loyal edge within the $r$-neighborhood of $u$.
From now on, our goal is to show that there exists some vertex $u$ with $|D(u,r)|$ significantly greater than $d^2$.

$\ $

Let $\alpha$ be a parameter to be determined later. The goal is to find a vertex $u$ such that $|D(u,r)|>\alpha d^2$.
Assume towards contradiction that \(|D(u,r)|\le \alpha d^2\) for every \(u\). The remainder of the proof consists of two steps: show that there is a large set of loyal edges, with every pair of them belonging to the same demand pair staying far from each other; and then show that there exists some small neighborhood that captures many edges in this set and therefore touches many distinct demand pairs.

\subsubsection*{Step 2: Finding a dispersed set of loyal edges}  
By similar arguments, we can first derive that every radius-\(r\) neighborhood contains $O(\alpha dr)$ vertices.
This is since (i) any demand contributes only \(O(r)\) loyal edges incident to a radius-\(r\) ball, and so the total number of loyal edges in such a ball is \(O(\alpha d^2r)\); and (ii) each vertex is incident to $d$ loyal edges.

This neighborhood size bound implies a bound on the number of local loyal edges.  Consider a vertex \(u\).  Let \(N_{\mathrm{loy}}(u)\) be the set of loyal neighbors of \(u\), and let \(P(u)\) be the set of demands that own a loyal edge incident to a vertex in \(N_{\mathrm{loy}}(u)\). 
By the Bodwin-Williams bound, handling these demands costs
\[
    \lambda(u,r)
    \le
    \Oh\!\left(
        \sqrt{|B(u,r)|\cdot |P(u)|\cdot |N_{\mathrm{loy}}(u)|}
    \right)
\]
edges.
Since \(|B(u,r)|\le O(\alpha dr)\), \(|P(u)|\le |D(u,r)|\le \alpha d^2\), and \(\sum_u |N_{\mathrm{loy}}(u)|=O(nd)\), summing over all vertices with Cauchy-Schwarz inequality, we get that
$\sum_u \lambda(u,r)\le \Oh(\alpha nd^2r^{1/2})$.

Let $E^{\mathrm{loy}}$ be the set of all loyal edges. For a loyal edge \(e\), define
\[
    \gamma(e,r)=\big|\{f\in E^{\mathrm{loy}}:
        \operatorname{dist}_H(e,f)\le r
        \text{ and $f,e$ belong to the same demand pair}
        \}\big|.
\]
Note that if loyal edges \(f,e\) of the same demand pair lie within distance \(r\) from each other, then \(e\) is forced to appear in the local preserver (that is, contributes $1$ to all $\lambda(u,r)$) for all vertices $u$ adjacent to an endpoint of $f$. Since all vertices are rich, each such \(f\) supplies \(\Omega(d)\) such vertices, and so
\[
    \sum_{e\in E^{\mathrm{loy}}}\gamma(e,r)
    \le
    \Oh(\alpha ndr^{1/2}).
\]
We now find a dispersed family of loyal edges.  For each demand \(p\), form a conflict graph on the loyal edges owned by \(p\), joining two such edges if their distance in \(H\) is at most \(r\).  
An independent set\footnote{In \Cref{sec:shared}, what we actually select is a set of $\Omega(nd)$ loyal edges such that each selected edge is close to \emph{few} other selected edges; for the purpose of this technical overview, we just assume that each selected edge is close to \emph{no} other selected edges, hence the ``independent set''.} in this graph is a set of loyal edges owned by \(p\) that are pairwise farther than \(r\).  Applying the Caro--Wei bound to each demand separately and then using Cauchy--Schwarz, we obtain a set \(I(r)\subseteq E^{\mathrm{loy}}\) such that no two edges of \(I(r)\) with the same owner are within distance \(r\), and
\[
    |I(r)|
    \ge
    \frac{|E^{\mathrm{loy}}|^2}
         {\sum_{e\in E^{\mathrm{loy}}}\gamma(e,r)}
    \ge
    \Omega\!\left(\frac{nd}{\alpha r^{1/2}}\right),
\]
using \(|E^{\mathrm{loy}}|=\Omega(nd)\).  

\subsubsection*{Step 3: Capturing many loyal edges in a neighborhood}
In graph $G$, choose centers greedily so that distinct centers are at least \(2r\) apart; assign all vertices in \(B(c,r)\) to \(c\), and assign every remaining vertex to a center within distance \(2r\), so every center hosts a cluster with diameter \(O(r)\).
Since every vertex has loyal degree $\Omega(d)$, $|B(c,r)|\ge \Omega(dr)$.
Since the radius-$r$ balls of all centers are disjoint, the number of clusters is at most $O\!\left(\frac{n}{dr}\right)$.

Consider now the separated set of edges computed in Step 2. By the pigeonhole principle, there must be some cluster \(B(c,r)\) that contains many selected loyal edges
\[
    |I(r)\cap B(c,r)|
    \ge
    \Omega\!\left(
        \frac{|I(r)|}{n/(dr)}
    \right)
    =
    \Omega\!\left(\frac{d^2r^{1/2}}{\alpha}\right).
\]
Since the edges in $I(r)\cap B(c,r)$ are within distance $O(r)$ from each other, from the property of set $I(r)$, each edge belongs to a distinct demand pair. Therefore, the center $c$ sees
$\Omega(\frac{d^2r^{1/2}}{\alpha})$
distinct demand pairs in its $r$-neighborhood.
Balancing this with the assumption that $|D(u,r)|\le \alpha d^2$ holds for all $u$ by taking \(\alpha=\Theta(r^{1/4})\), we get that there exists a vertex \(u\) with
\[|D(u,r)|\ge \Omega(d^2r^{1/4}).\] Although this is worse than the $\Omega(d^2 r)$ bound in the edge-disjoint case, we still manage to achieve the goal of discovering $\omega(d^2)$ demands in a small neighborhood, which is enough for an improvement over the Bodwin-Williams $np^{1/3}$ bound.

\section{Preliminaries}
\label{sec: prelim}

Let $G=(V,E)$ be an undirected unweighted simple graph with $n=|V|$.
For each node $u\in V$, let $N_G(u)$ be its set of neighbors, and let
$\dist_G(u,v)$ denote the shortest-path distance between $u$ and $v$.
For $U\subseteq V$, we use  $\diam_G(U)$ to denote the diameter
in $G$.

For $r\ge0$, define the radius-$r$ neighborhood
$B_G(u,r)=\{v\in V:\dist_G(u,v)\le r\}$; in particular,
$B_G(u,1)=N_G(u)\cup\{u\}$.
The radius-$r$ edge neighborhood is
$E_G(u,r)=\{e\in E:e\cap B_G(u,r)\ne\varnothing\}$.

Let $P$ be a set of unordered vertex pairs, called a \emph{demand set},
and write $p=|P|$. We assume demands are reachable between two distinct vertices. 

\paragraph{Minimal Distance Preservers.} We say a distance preserver $H=(V_H,E_H)$ of $(G,P)$ is \emph{minimal} if,
for every $e\in E_H$, the subgraph $H'=(V_H,E_H\setminus \{e\})$ by deleting $e$ is not a distance preserver of $(G,P)$. The minimality satisfies the following properties.

\begin{lemma}[\cite{CE06}]\label{lm: CE06}
Given an $n$-vertex graph $G$ and a set $P$ of $p$ demand pairs,
every minimal distance preserver of $(G,P)$ has
$O(n+p\sqrt n)$ edges.
\end{lemma}

\begin{lemma}[\cite{BVW21}]\label{lem: n + sqrt(n p sigma d)}
Let $S\subseteq V$ be nonempty, with $|S|=\sigma$ and
$\diam_G(S)\le d$, and let $P\subseteq S\times V$ be the demand set of size $|P|=p$.
For every minimal distance preserver $H$ of $(G,P)$,
\[
 |E_H|=O\!\left(n+\sqrt{np\sigma d}\right).
\]
\end{lemma}
We can further generalize \Cref{lem: n + sqrt(n p sigma d)} by a slight modification of the proof in \cite{BVW21}. For completeness, we give the proof in \Cref{appendix:remaining proofs}.
\begin{lemma}[Adapted from \cite{BVW21}]\label{lem: generalized p sigma d}
Let $S\subseteq V$ be nonempty, with $|S|=\sigma$ and bounded
$\diam_G(S)\le d$, and let $P\subseteq S\times V$ be the demand set of size $|P|=p$.
For every minimal distance preserver $H$ of $(G,P)$ and every
$U\subseteq V_H$ of size $k$, the subgraph $H[U]=(U, E_H^U)$ of $H$ induced by $U$ has edge size
\[
 |E_H^U|=O\!\left(k+\sqrt{kp\sigma d}\right).
\]
\end{lemma}

\paragraph{Essential Edges.}
An edge $e\in E_H$ is \emph{essential} to $q=(s,t)\in P$ in $H$ if
every shortest $s$--$t$ path in $H$ contains $e$. Define
\[
 \Pess_H^P(e)=\{q\in P:e\text{ is essential to }q\text{ in }H\}.
\]
We say that each demand in $\Pess_H^P(e)$ requires $e$, and write
$\Pess(e)$ when $H,P$ are clear. Minimality implies that every edge $e\in E_H$ has 
$\Pess_H^P(e)\ne\varnothing$. Otherwise, all demands
would retain shortest paths after deleting that edge.
A \emph{demand-label} $l(e)$ is any member of $\Pess_H^P(e)$.

\paragraph{Node Cover of Demands.}
Let $H$ be a minimal distance preserver of $(G,R)$.
A node $u\in V_H$ \emph{covers} a demand $q\in R$ if it is incident
to an edge $e$ essential to $q$, that is, $q\in\Pess_H^R(e)$.
In particular, every shortest path for $q$ in $H$ contains $u$.
A set $S\subseteq V_H$ covers $Q\subseteq R$ if, for every $q\in Q$,
some vertex of $S$ covers $q$.


\section{Warm-up: the Edge-Disjoint Case}
\label{sec: edge-disjoint}

In this section, we consider a minimal distance preserver in which
the demands can be connected by edge-disjoint shortest paths. We follow the same
approach as in the general case: find many demands covered by a small
set of nearby vertices, bound the number of edges needed for these
demands, and remove them. The edge-disjoint assumption makes both the
local counting argument and the demand-removal procedure simpler.

\begin{theorem}\label{thm:warm-up-simplified}
Let $G$ be an $n$-vertex undirected unweighted graph and let $P$ be a
set of $p$ demand pairs, where $n^{2/3}\le p\le n$. Let $H\subseteq G$
be a minimal distance preserver of $(G,P)$. Suppose that $H$ contains
one shortest path for each demand in $P$, with these paths mutually
edge-disjoint. Then
\[
                         |E(H)|=O(n^{5/6}p^{1/2}).
\]
\end{theorem}

Fix such paths and denote the path for a demand $q$ by $\pi(q)$.
For $Q\subseteq P$, write $\pi(Q)=\bigcup_{q\in Q}\pi(q)$ for their
union. In particular, $\pi(P)\subseteq H$ already preserves every
demand in $P$, so minimality of $H$ implies $H=\pi(P)$.
Throughout this section, retain all $n$ vertices, including isolated
vertices. All distances and balls are measured in $H$, including
when we select vertices from a subgraph of $H$.

\paragraph{Essential edges and demand deletion.}
Every edge of $\pi(q)$ is essential to $q$ in $H$. Indeed, minimality
makes it essential to some demand, while every other demand has a
chosen path avoiding that edge. Thus every shortest path for $q$ in
$H$ must contain every edge of $\pi(q)$, so $\pi(q)$ is unique. Its
subpaths are also unique shortest paths: replacing one by a different
shortest path would give a different shortest path for $q$.
Consequently, deleting a subset of demand paths leaves the other
paths intact and unique. Their union is still a minimal distance
preserver for the remaining demands. This lets us apply the same
arguments throughout the demand-removal procedure.

\paragraph{Preserving a covered set of demands.}
Recall that a vertex covers a demand if it is incident to an edge
essential to that demand. In our setting, a vertex covers a
nontrivial demand $q$ exactly when it lies on $\pi(q)$. Thus $S$
covers $Q$ if every path $\pi(q)$, $q\in Q$, meets $S$.
Trivial demands need no edges and can be omitted from any set we cover.

Suppose nonempty $S$ covers nonempty $Q$. For each $q=(s,t)\in Q$,
choose $u_q\in S\cap\pi(q)$ and split $q$ into $(u_q,s)$ and
$(u_q,t)$. Let $Q^*\subseteq S\times V_H$ be the resulting demand
set, omitting trivial and repeated pairs, so $|Q^*|\le2|Q|$.
Since $u_q$ lies on a shortest path for $q$,
\[
             \dist_H(s,u_q)+\dist_H(u_q,t)=\dist_H(s,t).
\]
Preserving the two split pairs therefore preserves $q$. Moreover,
their shortest paths in $H$ are unique subpaths of $\pi(q)$, so their
union is exactly $\pi(Q)$. In particular, $\pi(Q)$ is a minimal
distance preserver of $(H,Q^*)$.

Applying \Cref{lem: generalized p sigma d} with source set $S$ gives,
for every $U\subseteq V_H$,
\begin{equation}\label{eq:warm-up-local-bound}
 |E(\pi(Q)[U])|
 =O\!\left(|U|+
   \sqrt{|U|\,|Q|\,|S|\,\diam_H(S)}\right).
\end{equation}
If $S$ is a singleton, the unique paths from its vertex form a tree,
whose induced subgraph on $U$ has at most $|U|$ edges; hence the same
bound holds when $\diam_H(S)=0$. Taking $U=V_H$ bounds all the edges
of $\pi(Q)$, while taking $U$ to be a ball bounds its local edges.

We will find the covering set using the following lemma.

\begin{lemma}\label{lem:warm-up-cover}
Suppose $H$ has average degree at least $d$, every path $\pi(q)$ has
length at least $r$, and $1\le r\le d$. There are sets
$Q\subseteq P$ and $S\subseteq V_H$ such that $S$ covers $Q$ and
\[
 |Q|=\Omega(d^2r),\qquad
 |S|=O(|Q|/d),\qquad
 \diam_H(S)\le4r.
\]
\end{lemma}

The first bound ensures that each iteration removes many demands.
The other two bounds make \Cref{eq:warm-up-local-bound} small: the
number of covering vertices is at most $O(|Q|/d)$, and these vertices
are within distance $O(r)$ of one another. We first use the lemma to
prove the theorem, and then prove it by finding a dense subgraph,
growing neighborhoods, and selecting covering vertices greedily.

\subsection{Completing the Proof}

\begin{proof}[Proof of \Cref{thm:warm-up-simplified}]
Fix parameters $1\le r\le d$, whose values will be chosen at the end.
First remove all demand paths of length less than $r$. There are at
most $p$ such paths, so this removes at most $pr$ edges. Let $H'$ be
the union of the remaining paths, retaining all $n$ vertices.

While $H'$ has more than $nd/2$ edges, its average degree exceeds
$d$, and every remaining path has length at least $r$. We may
therefore apply \Cref{lem:warm-up-cover} to $H'$. It gives a demand
set $Q$ covered by $S$ with
$|Q|=\Omega(d^2r)$, $|S|=O(|Q|/d)$, and $\diam_{H'}(S)\le4r$.
Using \Cref{eq:warm-up-local-bound} in $H'$ with $U=V_H$, we get
\[
 |E(\pi(Q))|
 =O\!\left(n+\sqrt{n\,|Q|\,|S|\,\diam_{H'}(S)}\right)
 =O\!\left(n+|Q|\sqrt{\frac{nr}{d}}\right).
\]
Here the second bound follows by substituting $|S|=O(|Q|/d)$ and
$\diam_{H'}(S)=O(r)$. Remove $Q$ and its paths from $H'$. The
remaining paths are still edge-disjoint and unique, so their union
is again a minimal preserver and the procedure can continue.

Each iteration removes $\Omega(d^2r)$ demands, and no demand is
removed twice. There are therefore $O(p/(d^2r))$ iterations, so the
$O(n)$ term above contributes $O(np/(d^2r))$ edges in total. The
terms proportional to $|Q|$ sum to at most $O(p\sqrt{nr/d})$,
because the removed demand sets are disjoint subsets of $P$.
At termination, at most $nd/2$ edges remain. Including the short
paths removed initially, we have accounted for every edge of $H$:
\begin{equation}\label{eq:warm-up-total}
 |E(H)|
 =O\!\left(pr+\frac{np}{d^2r}
                  +p\sqrt{\frac{nr}{d}}+nd\right).
\end{equation}

To choose the parameters, first balance the two terms contributed by
the iterations. The equality
$np/(d^2r)=p\sqrt{nr/d}$ gives $dr=n^{1/3}$; substituting this relation
makes each term equal to $n^{2/3}p/d$. Balancing this with the final
$nd$ edges gives
\[
                 d=n^{-1/6}p^{1/2},\qquad r=n^{1/2}p^{-1/2}.
\]
These values satisfy $r\ge1$ because $p\le n$, and $r\le d$ because
$p\ge n^{2/3}$. Each of the last three terms in
\Cref{eq:warm-up-total} is now $n^{5/6}p^{1/2}$. The short-path term
is $pr=n^{1/2}p^{1/2}\le n^{5/6}p^{1/2}$, proving the theorem.
\end{proof}

\subsection{Growth of Local Neighborhoods}

The main use of edge-disjointness is the following growth bound.
Many demands pass through the neighbors of a vertex in a dense
subgraph. Each contributes a long subpath inside a larger ball, and
edge-disjointness lets us add their lengths. Comparing this edge
count with \Cref{eq:warm-up-local-bound} forces the ball to contain
many vertices.

\begin{lemma}\label{lem:warm-up-growth}
Suppose every path $\pi(q)$ has length at least $r$, where
$1\le r\le d$. Let $H_C=(V^*,E_C)\subseteq H$ be a subgraph of
minimum degree at least $d/4$. Then every $u\in V^*$ satisfies
\[
                         |B_H(u,r)|=\Omega(dr^2).
\]
\end{lemma}

\begin{proof}
If $r<2$, the at least $d/4$ neighbors of $u$ in $H_C$ already give
$|B_H(u,r)|=\Omega(d)=\Omega(dr^2)$. Assume $r\ge2$.
Choose $S\subseteq N_{H_C}(u)$ of size
$\lceil d/4\rceil=\Theta(d)$, and let $P'$ be the demands whose
paths meet $S$.

We first count these demands. Every vertex of $S$ has degree at
least $d/4$ in $H_C$. On the other hand, any shortest path visits at
most three neighbors of $u$: the subpath between its first and last
such vertices has length at most two, using the route through $u$.
At each of these vertices, the path uses at most two incident edges.
Thus a demand accounts for at most six of the incidences counted in
$\sum_{v\in S}\deg_{H_C}(v)$. Every core edge belongs to exactly one
chosen path, so
\[
 |P'|\ge\frac16\sum_{v\in S}\deg_{H_C}(v)=\Omega(d^2).
\]

Next fix $q\in P'$ and a vertex of $S$ on $\pi(q)$. Since the whole
path has length at least $r$, one of its two directions from this
vertex has at least $r/2$ edges. The first $\lfloor r/2\rfloor$
edges in that direction lie in $B_H(u,r)$: every vertex on this
subpath is at distance at most $1+\lfloor r/2\rfloor\le r$ from
$u$. This gives $\Omega(r)$ local edges for each demand in $P'$.
The paths are edge-disjoint, so these contributions add up to
$\Omega(|P'|r)$ edges.

Write $x=|B_H(u,r)|$. Since $S$ covers $P'$, $|S|=\Theta(d)$, and
$\diam_H(S)\le2$, \Cref{eq:warm-up-local-bound} gives
\[
 \Omega(|P'|r)
 \le |E(\pi(P')[B_H(u,r)])|
 \le O\!\left(x+\sqrt{x\,|P'|\,d}\right).
\]
At least one of the two terms in the upper bound must be
$\Omega(|P'|r)$. The first gives $x=\Omega(|P'|r)$; squaring the
inequality for the second gives $x=\Omega(|P'|r^2/d)$. Therefore
\[
 x=\Omega\!\left(\min\left\{|P'|r,\frac{|P'|r^2}{d}\right\}\right)
   =\Omega(dr^2),
\]
since $|P'|=\Omega(d^2)$ and $r\le d$.
\end{proof}

\subsection{Proof of the Covering Lemma}

\begin{proof}[Proof of \Cref{lem:warm-up-cover}]
\textbf{Find a dense subgraph.}
Starting from $H$, repeatedly delete a vertex of current degree less
than $d/4$. Charge each deleted edge to the vertex whose deletion
removes it. Each vertex is charged fewer than $d/4$ edges, so fewer
than $nd/4$ edges are deleted in total. Since $H$ initially has at
least $nd/2$ edges, the remaining subgraph $H_C=(V^*,E_C)$ satisfies
\[
 |E_C|\ge nd/4,\qquad \deg_{H_C}(v)\ge d/4\quad(v\in V^*).
\]
We use $H_C$ only to select vertices and edges. The demand paths and
all distances continue to be measured in $H$.

\textbf{Find a neighborhood containing many demands.}
Choose centers greedily from $V^*$, adding a new center whenever it
is at distance greater than $2r$ from every center already chosen.
By \Cref{lem:warm-up-growth}, the radius-$r$ ball around each center
contains $\Omega(dr^2)$ vertices. These balls are disjoint: a vertex
in two of them would give a path of length at most $2r$ between their
centers. There are consequently at most $O(n/(dr^2))$ centers.
When the selection ends, every vertex of $V^*$ is within distance
$2r$ of a center, since otherwise it could be added as another center.
Thus the radius-$2r$ balls cover $V^*$.

For $v\in V^*$, let $L(v)=\{q\in P:v\in\pi(q)\}$ be the demands
covered by $v$. A path uses at most two edges incident to $v$, giving
\[
 |L(v)|\ge\tfrac12\deg_{H_C}(v)\ge d/8,
 \qquad
 \sum_{v\in V^*}|L(v)|\ge |E_C|\ge nd/4.
\]
The second inequality follows by summing the first degree bound:
each edge of $H_C$ is counted at both endpoints. Summing $|L(v)|$
over the radius-$2r$ balls counts every vertex of $V^*$ at least
once. As there are $O(n/(dr^2))$ such balls, one of them accounts for
an $\Omega(dr^2/n)$ fraction of this total. Its sum is therefore at
least $\Omega(dr^2/n)\cdot(nd/4)=\Omega(d^2r^2)$.
Choose its center $u^*$ and put
$B^+=V^*\cap B_H(u^*,2r)$ and $\sigma=|B^+|$. Then
\begin{equation}\label{eq:warm-up-incidences}
 \Delta:=\sum_{v\in B^+}|L(v)|=\Omega(d^2r^2),
 \qquad \Delta\ge\sigma d/8.
\end{equation}
Here $\Delta$ counts pairs $(v,q)$ with $v\in B^+$ covering $q$.
A demand is counted at most $4r+1$ times: its first and last vertices
in $B^+$ have distance at most $4r$ via $u^*$, so the shortest subpath
between them has at most $4r+1$ vertices. This bounds all its visits
to $B^+$, even if the path leaves the ball between visits.

\textbf{Choose a small covering set.}
The preceding count guarantees many distinct demands in the ball.
We also need to cover many of them using few vertices. Start with
$S=Q=\varnothing$. Whenever a vertex $v\in B^+$ covers at least
$\Delta/(2\sigma)$ demands outside $Q$, add $v$ to $S$ and add
$L(v)$ to $Q$. Stop when no such vertex remains.

Every selected vertex adds at least $\Delta/(2\sigma)$ new demands.
Since these newly covered demands are counted only once,
\[
                 |S|\le\frac{2\sigma}{\Delta}|Q|
                      \le\frac{16}{d}|Q|,
\]
using the second bound in \Cref{eq:warm-up-incidences}. At
termination, each of the $\sigma$ vertices covers fewer than
$\Delta/(2\sigma)$ demands outside $Q$. Thus fewer than $\Delta/2$
of the original pairs $(v,q)$ involve uncovered demands, and at
least $\Delta/2$ involve demands in $Q$. Each demand is counted at
most $4r+1$ times, so
\[
 \frac{\Delta}{2}
 \le\sum_{v\in B^+}|L(v)\cap Q|
 \le(4r+1)|Q|.
\]
The first bound in \Cref{eq:warm-up-incidences} now gives
$|Q|=\Omega(d^2r)$. By construction, $S$ covers $Q$, and
$S\subseteq B_H(u^*,2r)$ gives $\diam_H(S)\le4r$.
\end{proof}

The selection of separated centers and the greedy covering step are
also used in the general case; \Cref{lem:packing,lem:cover} give
these arguments in the forms needed there. The additional work comes
from allowing paths to share edges. In the neighborhood-growth proof,
we used edge-disjointness to turn $|P'|$ long subpaths into
$\Omega(|P'|r)$ distinct edges. In the demand-removal procedure, we
used it to delete a demand's path without affecting the other paths.
The later sections replace these two uses by counting suitable
essential edges and bounding the edges lost when covered demands are
removed. The goal remains the same: cover many demands with few
nearby vertices and use the distance-preserver bound for the split
demands.

\section{Technical Lemmas}
\label{sec:shared}


Let $H=(V_H,E_H)$ be a minimal distance preserver of $(G,\RR)$, with
$|V_H|\le n$ and $|E_H|\le nd$. 
For an edge $e\in E_H$, define 
\[
 D(e)=\{q\in\RR:\text{every shortest path for $q$ in $H$ contains $e$}\}.
\]
Minimality implies $D(e)\ne\varnothing$. An edge is \emph{private} if
$|D(e)|=1$. Write $\Lambda_H=\{e\in E_H:|D(e)|=1\}$ and $\Lambda_H(q)=\{e\in\Lambda_H:D(e)=\{q\}\}$. The private length of $q$ is $\lambda_H(q)=|\Lambda_H(q)|$.
For a private edge, we write $D(e)=q$ to indicate $D(e)=\{q\}$. Assume that
\begin{equation}\label{eq:length-assumption}
                      \lambda_H(q)\ge l\quad\quad (q\in\RR).
\end{equation}

\begin{definition}[Demand Cover]\label{def:demand-cover}
Let $H$ be a minimal distance preserver of $(G,R)$. A vertex $x$
\emph{covers} a demand $q\in R$ if it is incident to an edge essential
to $q$ in $H$. A set $S\subseteq V_H$ covers $Q\subseteq R$ if every
demand in $Q$ is covered by a vertex of $S$.
\end{definition}

\begin{definition}[Demand Hitting]\label{def:demand-hitting}
Let $H$ be a minimal distance preserver of $(G,R)$. A vertex $u$
\emph{hits} a demand $q\in R$ if there are an edge $e$ essential to
$q$ in $H$ and an endpoint $x\in e$ such that $ux\in E_H$.
We call $(u,x,e)$ a \emph{hitting tuple} for $q$.
A set $S\subseteq V_H$ hits $Q\subseteq R$, or is a hitting set for
$Q$, if every demand in $Q$ has a hitting tuple whose first vertex
belongs to $S$.
\end{definition}
The technical hitting lemma below constructs these witnesses using
loyal edges. The definition also permits other essential edges, as
needed in \Cref{sec:plus2-pairwise}.

We introduce a parameter $r$ such that
\begin{equation}\label{eq:large-range}
                         6\le r\le\min\{d^2,l\},
\end{equation}
and we prove the following lemmas.

\begin{lemma}\label{lem:technical-exact-preserver}
Let $H=(V_H,E_H)$ be a minimal distance preserver of $(G,\RR)$ with
$|V_H|\le n$, $|E_H|\le nd$, and $|\Lambda_H|\ge nd/20$.
Suppose $\lambda_H(q)\ge l$ for every $q\in\RR$.
For every real parameter $ 6\le r\le\min\{d^2,l\},$
there are $Q\subseteq\RR$ and $S\subseteq V_H$ such that $S$ covers
$Q$ and
\[
 |Q|=\Omega(d^2\sqrt r),\qquad
 |S|=O(|Q|/d),\qquad \diam_H(S)=O(r).
\]
\end{lemma}

\begin{lemma}\label{lem:technical-plus2-hitting}
Let $H=(V_H,E_H)$ be a minimal distance preserver of $(G,\RR)$ with
$|V_H|\le n$ and $|E_H|\le nd$. Suppose it has a loyal core as defined
in \Cref{lem:core} below, satisfying \Cref{eq:core}, and
$\lambda_H(q)\ge l$ for every $q\in\RR$.
For every real parameter $ 6\le r\le\min\{d^2,l\},$
there are $Q\subseteq\RR$ and $S\subseteq V_H$ such that $S$ hits
$Q$ and
\[
 |Q|=\Omega(d^2\sqrt r),\qquad
 |S|=O(|Q|/d^2),\qquad \diam_H(S)=O(r).
\]
\end{lemma}

\paragraph{Overview.}
\Cref{lem:technical-exact-preserver,lem:technical-plus2-hitting} find a
large set of demands witnessed by a small set of nearby vertices. The
covering conclusion uses vertices incident to essential edges, while
the hitting conclusion permits one additional adjacency. We first
obtain many loyal edges whose labels have bounded local multiplicity.
We then cluster vertices in decreasing order of ball volume and use
\Cref{lem:cover} to select an efficient partial cover. \Cref{sec:proof-main}
will establish the private-length and density hypotheses needed to
apply \Cref{lem:technical-exact-preserver} to a changing residual graph.

\subsection{Basic Tools}
\label{subsec:technical-basic-tools}

\paragraph{Shortest Paths and Their Neighbors}
Fix an arbitrary shortest path $\pi(q)$ in $H$ for each $q\in\RR$. Every edge with $q\in D(e)$
lies on every such path. We start from the following results.

\begin{lemma}\label{lem:pathfacts}
For every demand $q\in \RR$:
\begin{enumerate}
\item A vertex is incident to at most two edges essential to $q$.
\item For every vertex $u$, at most three vertices of $\pi(q)$ belong to
$N_H(u)$. Therefore, there are at most six pairs $(x,e)$ where the edge $e\in E_H$ is essential to $q$ and $x\in N_H(u)$ is an endpoint of $e$.
\item If $e$ is essential to $q$ and a subpath $P$ of $\pi(q)$ contains
$e$, then every shortest path between the endpoints of $P$ contains $e$.
\end{enumerate}
\end{lemma}
\begin{proof}
For the first statement, every edge essential to $q$ lies on the
fixed simple path $\pi(q)$, which has at most two incident edges at a vertex. For the second statement, the first and last
neighbors of $u$ on the shortest path $\pi(q)$ have distance at most two. Thus, at most three vertices of $\pi(q)$ belong to
$N_H(u)$. By the first statement, each such vertex is incident to at most two edges essential to $q$, giving at most six pairs $(x,e)$. For the third statement, let $a,b$ be the two endpoints of the path $P$. Suppose there exists a shortest path between $a$ and $b$ avoiding
$e$. Then there exists a shortest path for $q$ avoiding
$e$. This leads to a contradiction that $e$ is essential to $q$.
\end{proof}

\paragraph{Dense Loyal Core.}
Suppose that $H=(V_H,E_H)$ satisfies $|\Lambda_H|\ge nd/20$.
The following pruning procedure gives a dense loyal core.
\begin{lemma}[Dense loyal core]\label{lem:core}
If $|\Lambda_H|\ge nd/20$, there is a subgraph $H_L=(V^*,\LL)$ of $H$ with nonempty vertex set $V^*\subseteq V_H$ and edge set $ \LL=\{\{u,v\}\in\Lambda_H:u,v\in V^*\}$
such that
\begin{equation}\label{eq:core}
 m:=|\LL|\ge\frac{nd}{40},\qquad
                    \deg_{H_L}(x)\ge\frac d{40}\quad(x\in V^*).
\end{equation}
\end{lemma}
\begin{proof}
Start with $(V_H,\Lambda_H)$ and repeatedly delete a vertex of current
degree less than $d/40$. Each edge is counted only when it is deleted, so
fewer than $nd/40$ edges are deleted in total. At least $nd/40$ edges
remain, and every remaining vertex has degree at least $d/40$.
\end{proof}
We call $H_L=(V^*,\LL)$ a dense loyal core of the graph $H$. Vertices in $V^*$ are \emph{rich}, and edges in $\LL$ are \emph{loyal}. In addition, given $a\ge 0$, a node $x\in V_H$ and demand $q\in \RR$, we define 
\begin{align*}
 B_H(x,a)&=\{v:\dist_H(x,v)\le a\}, &
 E_H(x,a)&=\{e\in E_H:e\cap B_H(x,a)\ne\varnothing\},\\
 B_H^*(x,a)&=B_H(x,a)\cap V^*,&
 \LL(x,a)&=E_H(x,a)\cap\LL,\\
 \LL(x,a,q)&=\{e\in\LL(x,a):D(e)=q\},&
 \lambda_H^*(x,a,q)&=|\LL(x,a,q)|.
\end{align*}

\paragraph{Growth of Local Balls.}
We prove a lower bound on the size of $B_H(x,a)$ for every rich vertex
$x\in V^*$.

\begin{lemma}\label{lem:growth}
Suppose $l\ge1$ and the loyal core has minimum degree at least
$\delta>0$. For every rich vertex $u$ and every real $a\ge1$,
\begin{equation}\label{eq:growth}
                    |B_H(u,a)|\ge\frac{\delta}{6}\min\{a,l\}.
\end{equation}
In particular, under \Cref{eq:core}, this is
$\Omega(d\min\{a,l\})$.
\end{lemma}
\begin{proof}
Fix a positive integer $j$ satisfying $2(j-1)<l$, and define
\[
 L_j=\{v\in V_H:\dist_H(u,v)=j\},\qquad
 W_j=V^*\cap B_H(u,j-1),\qquad b_j=|L_j|.
\]
The set $W_j$ is nonempty since $u\in W_j$. For each $w\in L_j$, fix
an arbitrary shortest-path tree $T_w$ rooted at $w$ in its component of
$H$. We claim
\begin{equation}\label{eq:tree-union}
             E(H_L[W_j])\subseteq\bigcup_{w\in L_j}E(T_w[W_j]).
\end{equation}
Indeed, take $e\in E(H_L[W_j])$ and let $q=D(e)$. The path $\pi(q)$
has length at least $l>2(j-1)$. If both endpoints of $q$ belonged to
$B_H(u,j-1)$, their distance would be at most $2(j-1)$ via $u$.
Hence at least one endpoint of $q$ is outside this ball. Orient the
traversal of $e$ along $\pi(q)$ toward such an endpoint, start at the
opposite endpoint of $e$, and stop at the first vertex $w$ of $L_j$.
The resulting subpath contains $e$. By \Cref{lem:pathfacts},
the tree path to $w$ also contains $e$. Thus $e\in T_w$, and both
endpoints of $e$ lie in $W_j$, proving \Cref{eq:tree-union}.

Each $T_w[W_j]$ is a forest. Therefore
\begin{equation}\label{eq:layer-upper}
                 |E(H_L[W_j])|\le b_j|W_j|.
\end{equation}
Every loyal neighbor of a vertex of $W_j$ that is not itself in $W_j$
belongs to $L_j$. Simplicity of the graph and the minimum degree
assumption give
\begin{equation}\label{eq:layer-lower}
             2|E(H_L[W_j])|\ge(\delta-b_j)|W_j|.
\end{equation}
Combining the last two inequalities and dividing by $|W_j|>0$ gives
$b_j\ge\delta/3$.

Now put $J=\min\{\lfloor a\rfloor,\lceil l/2\rceil\}$. For every
$1\le j\le J$, the inequality $2(j-1)<l$ holds. These $J$ layers are
disjoint and belong to $B_H(u,a)$. Since $a,l\ge1$,
$J\ge\tfrac12\min\{a,l\}$. Summing their sizes proves
\Cref{eq:growth}.
\end{proof}


\subsection{Loyal Edges with Bounded Conflict}
\label{subsec:technical-bounded-conflict}
In this subsection, fix a radius satisfying \Cref{eq:large-range},
and put \(n_x=|B_H(x,r)|\) for every rich vertex \(x\).
We prove the following lemmas.

\begin{lemma}[Good Loyal Edges]\label{lem:good}

Under \Cref{eq:large-range} and \Cref{eq:core}, there is a set
$\GG\subseteq\LL$, with $|\GG|\ge m/2=\Omega(nd)$, such that
for every $e\in\GG$ and every endpoint $x\in e$,
\begin{equation}\label{eq:good}
                  \lambda_H^*(x,r/2,D(e))
                        \le C\frac{n_x}{d\sqrt r},
\end{equation}
where $n_x=|B_H(x,r)|$ for every rich vertex $x$ and $C$ is an absolute constant.
\end{lemma}

\begin{lemma}[Localizing one demand label]\label{lem:localize}
Let $\GG$ be the good loyal edges supplied by \Cref{lem:good}, and fix
an absolute constant $A\ge0$. For $F\subseteq\GG$, choose an endpoint
$x_e$ of every $e\in F$. Suppose that
\[
 n_{x_e}\le N\quad(e\in F),\qquad
 \diam_H(\{x_e:e\in F\})\le Ar.
\]
Then every demand $q\in\RR$ satisfies
\begin{equation}\label{eq:label-bound}
 |\{e\in F:D(e)=q\}|=O\!\left(\frac{N}{d\sqrt r}\right).
\end{equation}
The implicit constant may depend on $A$.
\end{lemma}
To prove \Cref{lem:good,lem:localize}, we prove the following result.
\begin{lemma}
\label{lem:charge}
Under \Cref{eq:large-range} and \Cref{eq:core},
\begin{equation}\label{eq:charge}
 Z:=\sum_{x\in V^*}\ \sum_{\substack{e\in\LL\\x\in e}}
       \frac{\lambda_H^*(x,r/2,D(e))}{n_x}
       =O\!\left(\frac{m}{d\sqrt r}\right).
\end{equation}
\end{lemma}
\begin{proof}
Let $k=\lceil d/40\rceil$. Since \Cref{eq:large-range} implies
$d\ge\sqrt6$, we have $k=\Theta(d)$.
Partition $N_{H_L}(x)$ into bundles, each
of size between $k$ and $2k-1$. This is possible since
$\deg_{H_L}(x)\ge k$: form bundles of size $k$ and merge any nonempty final
remainder into the preceding bundle. Let $B_x(z)$ denote the bundle
containing the neighbor $z$ of $x$. Every bundle has size $\Theta(d)$.

For each incidence $(x,e)$ with $e=\{x,z\}\in\LL$, charge $1/n_x$
to every pair $(y,f)$ such that
\begin{equation}\label{eq:charging-rule}
       y\in B_x(z),\qquad f\in\LL(x,r/2,D(e)).
\end{equation}
The total charge is
\begin{equation}\label{eq:total-charge}
 T=\sum_{x\in V^*}\ \sum_{e=\{x,z\}\in\LL}
       \frac{|B_x(z)|\lambda_H^*(x,r/2,D(e))}{n_x}
       =\Theta(dZ).
\end{equation}

Fix a rich vertex $y$, and put
\[
 d_y=\deg_{H_L}(y),\qquad U_y=B_H(y,r/2+2),\qquad u_y=|U_y|.
\]
For each $x\in N_{H_L}(y)$ and each $z$ in the bundle of $x$
containing $y$, let $q=D(\{x,z\})=(s,t)$. Add $(x,s)$ and $(x,t)$
to a demand set $Q_y$; trivial pairs may be omitted. Repetitions are
removed. Then
\[
 Q_y\subseteq N_{H_L}(y)\times V_H,
                 \qquad |Q_y|\le4k d_y=O(d d_y).
\]
All vertices of $N_{H_L}(y)$ are at distance at most two from one
another in $H$. Choose a minimal distance preserver $H_y\subseteq H$
for $(H,Q_y)$. Applying \Cref{lem: generalized p sigma d} with source
set $N_{H_L}(y)$, $|Q_y|=O(dd_y)$, and vertex set $U_y$ gives
\begin{equation}\label{eq:local-size}
              |E(H_y[U_y])|=O(u_y+d_y\sqrt{d u_y}).
\end{equation}

If an incidence $(x,e)$ charges $(y,f)$, the two split pairs for
$q=D(e)=D(f)$ belong to $Q_y$. Since $e$ is essential to $q$, every
shortest path for $q$ passes through $x$, and
\[
                     \dist_H(s,x)+\dist_H(x,t)=\dist_H(s,t).
\]
Preserving the two split pairs thus preserves $q$. In particular
$f\in E(H_y)$ by essentiality. Both endpoints of $f$ lie within
distance $r/2+1$ of $x$, and $xy\in E_H$, so
\begin{equation}\label{eq:charged-edge-local}
                              f\in E(H_y[U_y]).
\end{equation}

Conversely, fix a pair $(y,f)$ that receives charge, and let
$q=D(f)$. Every charging incidence $(x,e)$ has $x\in N_H(y)$ and
$D(e)=q$. \Cref{lem:pathfacts} bounds the number of these
incidences by six. Moreover, since $r\ge6$,
\[
 U_y\subseteq B_H(x,r/2+3)\subseteq B_H(x,r),
                                  \qquad \frac1{n_x}\le\frac1{u_y}.
\]
Consequently
\[
                         T\le6\sum_{y\in V^*}\frac{|E(H_y[U_y])|}{u_y}.
\]
By \Cref{lem:growth}, since $r/2+2\ge r/2$ and $r\le l$,
we have $u_y=\Omega(dr)$. Combining this with
\Cref{eq:total-charge,eq:local-size} and $\sum_y d_y=2m$ gives
\begin{align}
 Z=O\!\left(\sum_{y\in V^*}
           \left(\frac1d+\frac{d_y}{\sqrt{d u_y}}\right)\right)
       \notag 
  =O\!\left(\frac nd+\frac{m}{d\sqrt r}\right).
    \label{eq:pre-absorption}
\end{align}
Finally, $m\ge nd/40$ and $\sqrt r\le d$ imply $\frac nd\le\frac{40m}{d^2}
                                  \le\frac{40m}{d\sqrt r}$.
This proves \Cref{eq:charge}.
\end{proof}
\paragraph{Completing the good-edge and localization proofs.}
\begin{proof}[Proof of \Cref{lem:good}]
Choose an absolute $C_0$ such that $Z\le C_0m/(d\sqrt r)$ in
\Cref{lem:charge}. Discard an edge if either endpoint contributes
more than $2C_0/(d\sqrt r)$ to the sum defining $Z$. Each discarded
edge accounts for a distinct term exceeding this threshold, so at most
$m/2$ edges are discarded. The remaining edges satisfy
\Cref{eq:good} with $C=2C_0$.
\end{proof}

\begin{proof}[Proof of \Cref{lem:localize}]
For a fixed $q$, the assertion is immediate if there are no relevant
edges. Otherwise every relevant endpoint $x_e$ belongs to $\pi(q)$.
The subpath from their first to their last occurrence has length at
most $Ar$, since a subpath of a shortest path is shortest. Partition
its vertices into consecutive blocks of at most
$\lfloor r/2\rfloor+1$ vertices. Each block has path diameter at most
$r/2$, and there are $O(A+1)$ blocks because $r\ge6$.

In every nonempty block of assigned endpoints choose one such endpoint
$x_i$, belonging to a good edge with label $q$. If $x_e$ lies in that
block, then $e\in\LL(x_i,r/2,q)$. \Cref{lem:good} therefore
bounds the number of these edges by $CN/(d\sqrt r)$ for each block.
Summing over the constant number of blocks proves the claim.
\end{proof}

\subsection{Greedy Clustering and Partial Set Cover}
\label{subsec:technical-clustering}

\begin{lemma}[Volume-decreasing clustering]\label{lem:packing}
Let $W\subseteq V_H$ and $a>0$. Put $v_x=|B_H(x,a)|$ for $x\in W$.
There are centers $\CC\subseteq W$ and a partition
$W=\bigsqcup_{c\in\CC}W_c$ such that
\[
 W_c\subseteq B_H(c,2a),\qquad
 v_x\le v_c\ (x\in W_c),\qquad
 \sum_{c\in\CC}v_c\le n,\qquad
 \sum_{c\in\CC}|W_c|=|W|\le n.
\]
\end{lemma}
\begin{proof}
Process $W$ in nonincreasing order of $v_x$, breaking ties arbitrarily.
Make $x$ a new center if its distance from every existing center is
greater than $2a$; otherwise assign it to any existing center at distance
at most $2a$. A center belongs to its own cluster. Each assigned center
was processed earlier, giving $v_x\le v_c$. Distinct centers have
distance greater than $2a$, so their radius-$a$ balls are disjoint.
The two summation bounds follow.
\end{proof}

\begin{lemma}[Partial set cover]
\label{lem:cover}
Let $\{A_i:i\in I\}$ be an indexed family of finite sets, let
$M=|I|\ge1$, and put
\[
 \Delta=\sum_{i\in I}|A_i|>0,\qquad
 \mu(q)=|\{i:q\in A_i\}|.
\]
If $\mu(q)\le\gamma$ for every element $q$, there are
$J\subseteq I$ and $Q=\bigcup_{i\in J}A_i$ such that
\begin{equation}\label{eq:partial-cover}
               |Q|\ge\frac{\Delta}{2\gamma},\qquad
               |J|\le\frac{2M}{\Delta}|Q|.
\end{equation}
\end{lemma}
\begin{proof}
Starting with no covered elements, repeatedly choose a set adding at
least $\Delta/(2M)$ new elements, until no such set exists. Each chosen
set adds at least that many new elements, proving the second inequality.
At termination every unchosen set contains fewer than $\Delta/(2M)$
uncovered elements, and every chosen set contains none. Hence
\[
 \sum_{i\in I}|A_i\setminus Q|\le\Delta/2,
 \qquad
 \sum_{i\in I}|A_i\cap Q|\ge\Delta/2.
\]
The latter sum is $\sum_{q\in Q}\mu(q)\le\gamma|Q|$, proving the
first inequality. Notice that the required multiplicity assumption is
an \emph{upper} bound, $\mu(q)\le\gamma$.
\end{proof}

\subsection{\texorpdfstring{Proof of \Cref{lem:technical-exact-preserver}}{Proof of the covering lemma}}
\label{subsec:technical-exact-proof}
\label{sec:26}

\begin{proof}
Construct the loyal core by \Cref{lem:core}. Take the good loyal edges $\GG$ from
\Cref{lem:good}.
Apply \Cref{lem:packing} to $W=V^*$ with $a=r$ and
$v_x=n_x=|B_H(x,r)|$. Denote the resulting vertex clusters by $V_c$.
Assign every good edge $e=\{x,y\}$ to one endpoint $x_e$ of maximum
radius-$r$ ball volume $n_{x_e}$, breaking ties arbitrarily. Put $e$ in $E_c$ if
$x_e\in V_c$. Thus
\[
 \GG=\bigsqcup_{c\in\CC}E_c,\qquad
 \sum_c|E_c|=|\GG|=\Omega(nd),\qquad
 \sum_c(n_c+|V_c|)\le2n.
\]
By averaging, some center $c$ satisfies
\begin{equation}\label{eq:26-dense}
                         |E_c|=\Omega\bigl(d(n_c+|V_c|)\bigr).
\end{equation}
Fix this center. Its assigned endpoints lie in $V_c\subseteq B_H(c,2r)$,
have diameter at most $4r$, and satisfy $n_{x_e}\le n_c$.
\Cref{lem:localize} yields
\begin{equation}\label{eq:26-gamma}
 \gamma_c:=\max_{q\in\RR}|\{e\in E_c:D(e)=q\}|
                           =O\!\left(\frac{n_c}{d\sqrt r}\right).
\end{equation}

For every $x\in V_c$ define the set
\[
                  A_x=\{q:\text{some }e\in E_c
                                   \text{ has }x\in e\text{ and }D(e)=q\}.
\]
Only edges of $E_c$ are used in these sets. Set $M=|V_c|$ and
$\Delta=\sum_{x\in V_c}|A_x|$. Count the incidences $(x,e)$ with
$x\in V_c$, $e\in E_c$, and $x\in e$. Each edge contributes at least
one incidence, through its assigned endpoint, and at most two.
Each vertex-label pair $(x,q)$ accounts for at most two incidences by
\Cref{lem:pathfacts}. Therefore
\begin{equation}\label{eq:26-incidences}
                          |E_c|/2\le\Delta\le2|E_c|.
\end{equation}
Moreover the multiplicity of any $q$ among the $A_x$ is at most
$2|\{e\in E_c:D(e)=q\}|\le2\gamma_c$.

Apply \Cref{lem:cover} with $\gamma=2\gamma_c$. It returns
$S\subseteq V_c$ and $Q=\bigcup_{x\in S}A_x$ satisfying
\begin{align*}
 |Q|&\ge\frac{\Delta}{4\gamma_c}
       =\Omega\!\left(\frac{|E_c|}{\gamma_c}\right)
       =\Omega\!\left(\frac{d n_c}{n_c/(d\sqrt r)}\right)
       =\Omega(d^2\sqrt r),\\
 |S|&\le\frac{2|V_c|}{\Delta}|Q|
       =O\!\left(\frac{|V_c|}{|E_c|}|Q|\right)
       =O(|Q|/d).
\end{align*}
Here we used \Cref{eq:26-dense,eq:26-gamma,eq:26-incidences}.
Every membership $q\in A_x$ witnesses that $x$ covers $q$, so $S$
covers $Q$. Finally, $S\subseteq B_H(c,2r)$ gives
$\diam_H(S)\le4r$.
\end{proof}

\subsection{\texorpdfstring{Proof of \Cref{lem:technical-plus2-hitting}}{Proof of the hitting lemma}}
\label{subsec:technical-hitting-proof}
\label{sec:40}

\begin{proof}
Assume $r\ge6$ and let
$\GG$ be the same good-edge set as in \Cref{lem:good}.
\paragraph{Selecting hitting tuples.}
Set $k=\lceil d/40\rceil=\Theta(d)$. For every rich vertex $x$, pick
an arbitrary set $A(x)\subseteq N_{H_L}(x)$ of exactly $k$ loyal
neighbors. Set up 
\[
 \mathcal T=\{(u,x,e):e\in\GG,\ x\in e,\ u\in A(x)\}.
\]
Every tuple witnesses that $u$ hits $D(e)$. Because both endpoints of
each good edge are rich,
\begin{equation}\label{eq:tuple-total}
                         |\mathcal T|=2k|\GG|=\Omega(nd^2).
\end{equation}
For a fixed pair $(u,q)$, by \Cref{lem:pathfacts}, at most six tuples have first coordinate $u$
and edge label $q$. 
\paragraph{Clustering the hitting vertices.}
Define
\[
 W=\{u:(u,x,e)\in\mathcal T\text{ for some }x,e\},
 \qquad N_u=|B_H(u,2r)|\quad(u\in W).
\]
Apply \Cref{lem:packing} to $W$ with radius $a=2r$. Write $V_c$
for its clusters, so that
\[
 V_c\subseteq B_H(c,4r),\qquad N_u\le N_c\ (u\in V_c),\qquad
                       \sum_c(N_c+|V_c|)\le2n.
\]
Assign each tuple to the cluster of its first coordinate:
$\mathcal T_c=\{(u,x,e)\in\mathcal T:u\in V_c\}$.
By \Cref{eq:tuple-total} and weighted averaging, some center $c$ satisfies
\begin{equation}\label{eq:40-dense}
                       |\mathcal T_c|
                           =\Omega\bigl(d^2(N_c+|V_c|)\bigr).
\end{equation}
Fix this center $c$ for the rest of the proof.

\paragraph{Bounding repetitions of a demand.}
Let $F_c$ be the set of distinct edges appearing in $\mathcal T_c$.
For each $e\in F_c$ choose one witnessing tuple $(u_e,x_e,e)$.
Since $u_e\in V_c$ and $u_ex_e\in E_H$,
\[
 x_e\in B_H(c,4r+1),\qquad
 B_H(x_e,r)\subseteq B_H(u_e,r+1)\subseteq B_H(u_e,2r).
\]
Consequently $n_{x_e}\le N_{u_e}\le N_c$, and the assigned endpoints
have diameter at most $8r+2\le9r$. Apply
\Cref{lem:localize} to $F_c$. Therefore, for every $q$,
\[
                         |\{e\in F_c:D(e)=q\}|
                                =O\!\left(\frac{N_c}{d\sqrt r}\right).
\]
Each distinct edge contributes at most $2k=O(d)$ selected tuples.
It follows that
\begin{equation}\label{eq:40-gamma}
 |\{(u,x,e)\in\mathcal T_c:D(e)=q\}|
                            =O\!\left(\frac{N_c}{\sqrt r}\right).
\end{equation}

\paragraph{Partial set cover.}
For each $u\in V_c$, define $ A_u=\{q:\text{there is a tuple }(u,x,e)\in\mathcal T_c
          \text{ with }D(e)=q\}.$
Put $M=|V_c|$ and $\Delta=\sum_{u\in V_c}|A_u|$. Each incidence
$(u,q)$ is witnessed by at least one and at most six tuples, hence
\begin{equation}\label{eq:40-incidences}
                       |\mathcal T_c|/6\le\Delta\le|\mathcal T_c|.
\end{equation}
By \Cref{eq:40-gamma}, each demand occurs in at most
$\gamma=O(N_c/\sqrt r)$ of the sets $A_u$. Applying \Cref{lem:cover}
gives $S\subseteq V_c$ and $Q=\bigcup_{u\in S}A_u$ with
\begin{align*}
 |Q|&\ge\frac{\Delta}{2\gamma}
       =\Omega\!\left(\frac{|\mathcal T_c|}{N_c/\sqrt r}\right)
       =\Omega(d^2\sqrt r),\\
 |S|&\le\frac{2|V_c|}{\Delta}|Q|
       =O\!\left(\frac{|V_c|}{|\mathcal T_c|}|Q|\right)
       =O(|Q|/d^2),
\end{align*}
using \Cref{eq:40-dense} and \Cref{eq:40-incidences}. By the
definition of $A_u$, the set $S$ hits $Q$. Finally,
$S\subseteq B_H(c,4r)$ implies $\diam_H(S)\le8r$.
\end{proof}

\section{\texorpdfstring{Proof of \Cref{thm: main}}{Proof of the main theorem}}
\label{sec:proof-main}
\label{sec: general}
\label{sec:7-corrected}

In this section, we use \Cref{lem:technical-exact-preserver} in an
iterative demand-removal procedure to prove \Cref{thm: main}.

There are two types of deletion. A demand with few private edges can
be removed at small edge cost. Once no such demand remains, private
lengths are large, and their sum guarantees enough private edges to
apply \Cref{lem:technical-exact-preserver}. Its covered batch has a
small distance preserver, which bounds the edge loss when the batch is
removed. We continue until at most half the demands remain, while
retaining at least three quarters of the original edges. The induction
hypothesis then gives a contradiction.

We prove the following stronger statement: there is an absolute constant
$C$ such that, for every $n$-vertex graph $G$ and every demand set $P$
with $n^{1/4}\le p:=|P|\le n$, every minimal distance preserver
$H=(V_H,E_H)$ of $(G,P)$ satisfies
\begin{equation}\label{eq:s7-target}
                         |E_H|\le C n^{8/9}p^{4/9}.
\end{equation}
This implies \Cref{thm: main} in its stated range $n^{1/2}\le p\le n$.
We retain all $n$ vertices in every subgraph, including isolated
vertices. Trivial demands and demands between different connected
components may be discarded without changing which subgraphs preserve
the demands or which preservers are minimal. If fewer than $n^{1/4}$
nontrivial reachable demands remain, \Cref{lm: CE06} gives $O(n)$
edges, within the claimed bound for the original $p\ge n^{1/4}$.
Otherwise we prove the statement for the remaining demands, and the
original bound follows by monotonicity in $p$.

\subsection{Edge Removal Under Demand Deletion}
\label{subsec:main-deletion-loss}
We first discuss how removing demands affects a minimal distance preserver.

\begin{observation}[Removing one demand]\label{obs:s7-one-demand}
Let $K$ be a minimal distance preserver of $(K,R)$, let $q\in R$,
and put $R'=R\setminus\{q\}$. Every distance preserver $K'\subseteq K$
of $(K,R')$ contains every edge $e\in E(K)$ for which
$D_K^R(e)\ne\{q\}$. Consequently,
\begin{equation}\label{eq:s7-one-demand}
                    |E(K)|-|E(K')|\le\lambda_K(q).
\end{equation}
Here $\lambda_K(q)$ is the private length before $q$ is removed.
\end{observation}

\begin{proof}
Minimality gives $D_K^R(e)\ne\varnothing$ for every $e\in E(K)$.
If $D_K^R(e)\ne\{q\}$, choose a demand $q'\in D_K^R(e)\setminus\{q\}$.
Every shortest path for $q'$ in $K$ contains $e$. Since $K'$ preserves
$q'$, it contains such a path and hence contains $e$. Therefore, only edges
in $\Lambda_K(q)$ can be removed.
\end{proof}

\begin{lemma}[Removing a covered demand set]\label{lem:s7-covered-deletion}
Let $K$ be a minimal distance preserver of $(K,R)$, with at most $n$
vertices. Suppose that a nonempty set $S\subseteq V(K)$ covers a
nonempty demand set $Q\subseteq R$, and $\diam_K(S)<\infty$.
There is a distance preserver $H(Q)\subseteq K$ of $(K,Q)$ such that,
for every distance preserver $K'\subseteq K$ of $(K,R\setminus Q)$,
\begin{equation}\label{eq:s7-cover-cost}
 |E(K)|-|E(K')|\le |E(H(Q))|
 \le A\left(n+\sqrt{n|Q||S|\diam_K(S)}\right),
\end{equation}
where $A$ is an absolute constant.
\end{lemma}

\begin{proof}
For each $q=(s,t)\in Q$, choose a covering vertex $u_q\in S$.
By \Cref{def:demand-cover}, it is incident to an edge essential to
$q$, so it belongs to every shortest path for $q$ in $K$. Hence
\begin{equation}\label{eq:s7-split-distance}
 \dist_K(s,u_q)+\dist_K(u_q,t)=\dist_K(s,t).
\end{equation}
Form the split demand set
\[
 Q^*=\bigcup_{q=(s,t)\in Q}\{(u_q,s),(u_q,t)\}
       \subseteq S\times V(K),
\]
omitting repeated and trivial pairs. Thus $|Q^*|\le2|Q|$.
If $|S|=1$, choose a BFS tree rooted at its unique vertex, then a
minimal subgraph of that tree preserving $Q^*$. Otherwise choose a
minimal distance preserver of $(K,Q^*)$ and apply
\Cref{lem: generalized p sigma d}. In either case,
call the resulting subgraph $H(Q)$. It satisfies
\[
 |E(H(Q))|=O\!\left(n+\sqrt{n|Q||S|\diam_K(S)}\right).
\]
The two preserved routes in \Cref{eq:s7-split-distance} show that
$H(Q)$ is also a distance preserver of $(K,Q)$.

For any remaining-demand preserver $K'$, the subgraph $K'\cup H(Q)$
preserves all demands in $R$ exactly. Minimality of $K$ therefore gives
\begin{equation}\label{eq:s7-union}
 E(K')\cup E(H(Q))=E(K).
\end{equation}
Indeed, a missing edge would be essential to some demand in $R$,
contradicting preservation of that demand. Thus
$E(K)\setminus E(K')\subseteq E(H(Q))$, proving
\Cref{eq:s7-cover-cost} after fixing $A$ sufficiently large.
\end{proof}

\subsection{Induction and Fixed Parameters}
\label{subsec:main-induction}

Let $C_{\rm CE}$ be an absolute constant such that, by \Cref{lm: CE06}, every minimal distance preserver $J$ of an $n$-vertex graph
with demand set $T$ has size
\begin{equation}\label{eq:s7-ce}
              |E(J)|\le C_{\rm CE}\bigl(n+n^{1/2}|T|\bigr).
\end{equation}

We prove \Cref{eq:s7-target} by strong induction on the integer $p$,
uniformly over all $n$-vertex graphs and demand sets of size $p$ in
$[n^{1/4},n]$. 
\paragraph{Base Case} Consider the range $n^{1/4}\le p\le n^{7/10}$. It satisfies that $n\le n^{8/9}p^{4/9}$ and $n^{1/2}p\le n^{8/9}p^{4/9}$. By \Cref{eq:s7-ce} with $C\ge2C_{\rm CE}$, every minimal distance preserver $H=(V_H,E_H)$ of $(G,P)$ has size
\[
        |E_H|\le C_{\rm CE}\bigl(n+n^{1/2}p\bigr) \le C n^{8/9}p^{4/9}.
\]

\paragraph{Inductive Step} Consider the range $n^{7/10}<p\le n$. Assume the inductive statement for all smaller integer demand counts in $[n^{1/4},p-1]$.
We need the following result for small residual sets.

\begin{claim}\label{clm:s7-half}
Under the induction hypothesis, every minimal distance preserver $J$
of $(G,R)$ with $|R|\le p/2$ satisfies
\begin{equation}\label{eq:s7-half}
                       |E(J)|\le Cn^{8/9}(p/2)^{4/9}.
\end{equation}
\end{claim}
\begin{proof}
Set $p'=|R|$. Suppose $p'\ge n^{1/4}$. Given $p'<p$, by induction, $|E(J)|\le Cn^{8/9}(p')^{4/9} \le Cn^{8/9}(p/2)^{4/9}.$
Suppose $p'<n^{1/4}$. By \Cref{eq:s7-ce}, 
\[
|E(J)|\le C_{\rm CE}(n+n^{1/2}p')
\le C_{\rm CE}(n+n^{3/4})\le2C_{\rm CE}n.
\]
Since $p\ge n^{1/4}$, we have
$n^{8/9}(p/2)^{4/9}\ge2^{-4/9}n$. Choosing
$C\ge2^{13/9}C_{\rm CE}$ therefore gives the required bound
\[
 |E(J)|\le2C_{\rm CE}n\le Cn^{8/9}(p/2)^{4/9}.
\]
\end{proof}
We prove the inductive step by contradiction.
Suppose that a minimal distance preserver
$H=(V_H,E_H)$ of $(G,P)$ violates \Cref{eq:s7-target}. Fix
\begin{equation}\label{eq:s7-contradiction}
 n=|V_H|,\qquad d=|E_H|/n,\qquad
              nd=|E_H|>Cn^{8/9}p^{4/9}.
\end{equation}
The average degree of $H$ is $2d$. In addition, we define
\begin{equation}\label{eq:s7-parameters}
                  l=\frac{nd}{5p},\qquad
                  r_0=n^{1/2}d^{-3/2},\qquad
                  r=\max\{6,r_0\}.
\end{equation}
All five parameters $n,d,p,l,r$ remain fixed throughout the procedure.

\subsection{Demand Removal Procedure}
\label{subsec:main-demand-removal}

\paragraph{The residual graph and its update.}
At every stage, we use $R\subseteq P$ to denote the demands still present. Let $H(R)\subseteq H$ be the current minimal distance preserver of $(G, R)$. If the demand set changes from $R$ to $R'\subseteq R$,
pick an arbitrary minimal subgraph $H(R')$ of the current $H(R)$ that preserves all the demands in $R'$. 

Note that such an update exists: repeatedly remove one edge from $H(R)$ whose removal preserves the demands in $R'$, recomputing essentiality after
each deletion, until no further deletion is possible. If $K=H(R)$ is the graph just before the update and
$K'=H(R')$ the graph after it, then for every $q\in R'$,
\[
    \dist_{K'}(q)=\dist_K(q)=\dist_H(q)=\dist_G(q).
\]
Here $\dist_K(q)$ denotes the distance between the endpoints of $q$.
Thus $K'$ is a minimal distance preserver of $(G,R')$ as well as
of $(K,R')$. Essentiality and private lengths are recomputed in
each updated instance.
\paragraph{Demand-removal procedure.}
Set $R=P$ and $H(R)=H$ as the initialization. Repeat the following two steps while $|R|>p/2$.
\begin{enumerate}[label=\arabic*.,ref=\arabic*,leftmargin=*]
\item\label{step:main-single-removal}
\textbf{Remove demands of small private length.}
While $|R|>p/2$ and there exists a demand $q\in R$ with
$\lambda_{H(R)}(q)\le l$, remove $q$ from $R$, update $H(R)$ and recompute the private lengths. If $|R|\le p/2$, terminate the procedure.

\item\label{step:main-batch-removal}
\textbf{Remove a covered batch.}
Now $|R|>p/2$ and $\lambda_{H(R)}(q)>l$ for every $q\in R$.
The private-edge sets of different demands are disjoint, so
\begin{equation}\label{eq:s7-private-density}
 |\Lambda_{H(R)}|=\sum_{q\in R}\lambda_{H(R)}(q)
       >|R|l>\frac p2\frac{nd}{5p}=\frac{nd}{10}.
\end{equation}
Apply \Cref{lem:technical-exact-preserver} to the current graph
$K=H(R)$, with the fixed parameters $n,d,l,r$ defined in  \Cref{eq:s7-parameters} and verified in \Cref{lem:s7-batch}. 
We obtain nonempty sets
$Q\subseteq R$ and $S\subseteq V_H$ such that $S$ covers $Q$ in $K$,
and
\begin{equation}\label{eq:s7-group-properties}
 |Q|\ge c_0d^2\sqrt r,\qquad
 |S|\le c_1|Q|/d,\qquad \diam_K(S)\le c_2r,
\end{equation}
where $c_0>0,c_1,c_2$ are absolute constants from
\Cref{lem:technical-exact-preserver}. Remove $Q$ from $R$ and update
$H(R)$ to a minimal subgraph preserving all remaining distances.
\end{enumerate}

Each pass removes at least one demand, and no demand is reintroduced,
so the procedure terminates with $|R|\le p/2$.
All hypotheses of \Cref{lem:technical-exact-preserver} hold in the
current instance: $H(R)$ is a minimal distance preserver of $(G,R)$,
retains $n$ vertices, and has at most $nd$ edges by containment in $H$.
\Cref{eq:s7-private-density} gives the required private-edge mass,
and every active demand has private length greater than $l$.
The radius condition is checked below; the diameter conclusion is
measured in the current graph $H(R)$ itself.

\begin{lemma}[Total loss in \Cref{step:main-single-removal}]\label{lem:s7-single-total}
The total number of edges removed in \Cref{step:main-single-removal} is at most $nd/5$.
\end{lemma}
\begin{proof}
Immediately before a single demand $q$ is removed in \Cref{step:main-single-removal}, its
private length is at most $l$. \Cref{obs:s7-one-demand},
applied to that current graph and demand set, bounds the number of
edges lost in the update by $l$. There are at most $p$ such removals.
The total is therefore at most $pl=nd/5$.
\end{proof}

\begin{lemma}\label{lem:s7-batch}
For sufficiently large $C$, the parameters in
\Cref{eq:s7-parameters} satisfy $6\le r\le\min\{d^2,l\}$.
In each execution of \Cref{step:main-batch-removal}, removing its demand set $Q$ and then
updating the graph deletes at most
\begin{equation}\label{eq:s7-per-group}
                C_2\left(n+|Q|\sqrt{\frac{nr}{d}}\right)
\end{equation}
edges for an absolute constant $C_2$ independent of $C$.
\end{lemma}
\begin{proof}
Since $p>n^{7/10}$, \Cref{eq:s7-contradiction} implies $ d>Cn^{-1/9}p^{4/9}>Cn^{1/5}\ge C.$
Therefore
\[
          \frac{r_0}{d}=\frac{n^{1/2}}{d^{5/2}}<C^{-5/2},
                    \qquad l=\frac{nd}{5p}\ge\frac d5.
\]
If $C\ge30$, then $d>C\ge30$, so $6\le d/5$. Also
$C^{-5/2}\le1/5$, hence $r_0\le d/5$. Therefore
\[
 6\le r=\max\{6,r_0\}\le d/5\le\min\{d^2,l\},
\]
which verifies the radius condition of
\Cref{lem:technical-exact-preserver}.

Let $K=H(R)$ immediately before \Cref{step:main-batch-removal}. By
\Cref{lem:s7-covered-deletion,eq:s7-group-properties}, the number of edges deleted from $K$ is at most
\begin{align*}
 |E(H(Q))|
 &\le A\left(n+\sqrt{n|Q||S|\diam_K(S)}\right)\\
 &\le A\left(n+\sqrt{c_1c_2}\,|Q|\sqrt{\frac{nr}{d}}\right)
 \le C_2\left(n+|Q|\sqrt{\frac{nr}{d}}\right).
\end{align*}
\end{proof}

\begin{lemma}[Total loss in \Cref{step:main-batch-removal}]\label{lem:s7-group-total}
For sufficiently large $C$, the total number of edges removed in
\Cref{step:main-batch-removal} is at most $nd/20$.
\end{lemma}
\begin{proof}
Let $Q_i,S_i$, for $i=1,\ldots,\tau$, be the sets selected in the
executions of \Cref{step:main-batch-removal}. These demand sets are pairwise disjoint
subsets of $P$, so
\[
              \sum_{i=1}^{\tau}|Q_i|\le p,\qquad
              \tau\le\frac{p}{c_0d^2\sqrt r},
\]
by \Cref{eq:s7-group-properties}. Summing
\Cref{eq:s7-per-group}, the total loss is at most
\begin{equation}\label{eq:s7-total-before-parameters}
 C_1\left(\frac{np}{d^2\sqrt r}
                    +p\sqrt{\frac{nr}{d}}\right),
\end{equation}
where $C_1$ is an absolute constant determined only by $C_2$ and $c_0$.
This bound also holds when $\tau=0$.
We estimate the two terms with the actual radius
$r=\max\{6,r_0\}$. First, since $r\ge r_0$,
\begin{equation}\label{eq:s7-first-term}
 \frac{1}{nd}\frac{np}{d^2\sqrt r}
       \le\frac{p n^{-1/4}}{d^{9/4}}
       \le C^{-9/4}.
\end{equation}
The last inequality follows since 
$d^{9/4}>(Cn^{-1/9}p^{4/9})^{9/4}$.
For the second term, by the definition of $r$,
\begin{align}
 \frac{p}{nd}\sqrt{\frac{nr}{d}}
 &=\max\left\{
       \frac{p n^{-1/4}}{d^{9/4}},
       \frac{\sqrt6\,p}{n^{1/2}d^{3/2}}
             \right\}\notag\\
 &\le\max\left\{C^{-9/4},
       \sqrt6\,C^{-3/2}(p/n)^{1/3}\right\}
 \le\sqrt6\,C^{-3/2}.\label{eq:s7-second-term}
\end{align}
Here the second expression uses
$d^{3/2}>C^{3/2}n^{-1/6}p^{2/3}$, and the final inequality uses
$p\le n$ and $C\ge1$.
Combining \Cref{eq:s7-total-before-parameters,eq:s7-first-term,eq:s7-second-term}
bounds the total loss by
\[
                 C_1\bigl(C^{-9/4}+\sqrt6\,C^{-3/2}\bigr)nd.
\]
Choose $C$ large enough that the coefficient is at most $1/20$.
\end{proof}

\subsection{Closing the Induction}
\label{subsec:main-closing}
Fix $C$ large enough that
\[
 C\ge\max\{30,2^{13/9}C_{\rm CE}\},\qquad
 C_1\bigl(C^{-9/4}+\sqrt6\,C^{-3/2}\bigr)\le1/20.
\]
This choice satisfies all preceding requirements. There is no circular
dependence: $C_1$ depends only on the constants in the preliminary and
technical lemmas, not on $C$.

Let $R$ be the demand set at termination. By
\Cref{lem:s7-single-total,lem:s7-group-total}, at most $ \frac{nd}{5}+\frac{nd}{20}=\frac{nd}{4}$ edges have been deleted. Thus the residual graph satisfies
\begin{equation}\label{eq:s7-residual-lower}
|E(H(R))|\ge\frac34nd.
\end{equation}
It is a minimal distance preserver of $(G,R)$ and $|R|\le p/2$. By
\Cref{clm:s7-half},
\begin{equation}\label{eq:s7-residual-upper}
|E(H(R))|\le Cn^{8/9}(p/2)^{4/9}
=2^{-4/9}Cn^{8/9}p^{4/9}.
\end{equation}
Combining \Cref{eq:s7-residual-lower,eq:s7-residual-upper}, 
\[
nd\le\frac43\,2^{-4/9}Cn^{8/9}p^{4/9}
<Cn^{8/9}p^{4/9},
\]
contradicting \Cref{eq:s7-contradiction}. The induction proves
\Cref{eq:s7-target}, and therefore proves \Cref{thm: main}.\qed

\section{\texorpdfstring{Pairwise Additive \(+2\) Spanners}{Pairwise Additive +2 Spanners}}
\label{sec:plus2-pairwise}

We apply the framework for minimal distance preservers to additive
approximation. We keep the notation of \Cref{sec:shared,sec:proof-main}:
$R\subseteq P$ is the current demand set, $H(R)$ is a minimal distance
preserver of $(G,R)$, $Q\subseteq R$ is a selected batch, and $S$ is a
set of vertices that covers or hits $Q$. We use $J$ for the final
additive spanner and $J(Q)$ for the subgraph saved for a selected batch.

\begin{definition}[Pairwise Additive $+k$ Spanner]
\label{def:pairwise-additive-spanner}
Given an integer $k\ge0$, an $n$-vertex undirected unweighted graph
$G=(V,E)$, and a demand set $P\subseteq V\times V$, a subgraph
$J\subseteq G$ is a pairwise additive $+k$ spanner of $(G,P)$ if
\[
 \dist_G(s,t)\le\dist_J(s,t)\le\dist_G(s,t)+k
       \qquad((s,t)\in P).
\]
\end{definition}

\begin{theorem}\label{thm: main +2 spanner}
Let $G$ be any $n$-vertex undirected unweighted graph and let $P$ be
any set of $p$ demand pairs. Then $(G,P)$ admits a pairwise additive
$+2$ spanner with $O\!\left(n+n^{9/10}p^{2/5}\right)$ edges.
\end{theorem}

\paragraph{Proof strategy and comparison with the exact bound.}
Both \Cref{thm: main,thm: main +2 spanner} use minimal distance
preservers and the extraction lemmas of \Cref{sec:shared}. In each
case, the goal is to handle a batch of $\Omega(d^2\sqrt r)$ demands
using vertices of diameter $O(r)$. The difference is the witness set.
For \Cref{thm: main}, a covering vertex lies on a shortest path, and
\Cref{lem:technical-exact-preserver} gives $|S|=O(|Q|/d)$.
For the additive bound, a hitting vertex is adjacent to a shortest
path, and \Cref{lem:technical-plus2-hitting} gives the smaller set
$|S|=O(|Q|/d^2)$. Routing through a hitting vertex costs at most two
extra edges. After splitting each demand at its chosen vertex of $S$,
the respective bounds are
\begin{align*}
 |E(H(Q))|&=O\!\left(n+|Q|\sqrt{\frac{nr}{d}}\right)
             &&\text{for a covering set},\\
 |E(J(Q))|&=O\!\left(n+\frac{|Q|\sqrt{nr}}d\right)
             &&\text{for a hitting set}.
\end{align*}
Thus allowing additive error two saves a factor $\sqrt d$ in the
nonlinear batch cost.

There is also a difference in how these bounds are used.
\Cref{sec:proof-main} bounds every minimal distance preserver by an
edge-count contradiction. Its exact batch preserver $H(Q)$ bounds the
number of edges lost after deleting $Q$, because minimality forces
\Cref{eq:s7-union}. A $+2$ spanner for $Q$ does not give that union
identity. Instead, we construct $J$ explicitly: save each $J(Q)$ in
$J$, remove $Q$, and keep $H(R)$ a minimal distance preserver for the
remaining demands. The saved routes are compared with $\dist_G$, so
taking their union does not accumulate additive error.

This explains the different treatment of short demands. The
proof of \Cref{thm: main} deletes demands of small private length. Here we first purchase
shortest paths for demands of small original distance, at cost $O(pr)$.
The remaining minimal distance preserver need not satisfy the
private-edge hypotheses of \Cref{lem:technical-plus2-hitting}.
\Cref{lem:plus2-extraction} records the short extension of its proof
that uses assigned essential demands and long full distances. It yields
hitting sets in exactly the sense of \Cref{def:demand-hitting}, without
requiring a second proof of the extraction argument.

The batch sizes remain $\Omega(d^2\sqrt r)$, so summing the additive
batch costs gives
\[
 O\!\left(\frac{np}{d^2\sqrt r}+\frac{p\sqrt{nr}}d\right).
\]
Balancing these terms gives $r\approx\sqrt n/d$, compared with
$r\approx\sqrt n\,d^{-3/2}$ for \Cref{thm: main}. Including the final
$O(nd)$ minimal distance preserver gives $d\approx n^{-1/10}p^{2/5}$.
We check the radius bounds, short-path cost, and endpoint ranges below.


\subsection{Hitting Sets and Additive +2 Error}
\label{subsec:plus2-tools}

Let $H$ be a minimal distance preserver, let $S\subseteq V_H$ have
finite diameter $D$, and let $Q^*\subseteq S\times V_H$ be auxiliary
demands. The same estimates used in \Cref{lem:s7-covered-deletion},
namely \Cref{lem: generalized p sigma d}, give a
minimal distance preserver $H(Q^*)\subseteq H$ of $(H,Q^*)$ with
\begin{equation}\label{eq:plus2-source}
 |E(H(Q^*))|=O\!\left(n+\sqrt{n|Q^*||S|D}\right).
\end{equation}
For a singleton $S$, choose a BFS tree rooted at its vertex and then
a minimal subgraph preserving the auxiliary demands. This gives
$O(n)$ edges also when $D=0$.

\begin{lemma}[A hitting set gives an additive spanner]
\label{lem:plus2-hitting-repair}
Let $H$ be a minimal distance preserver of $(G,R)$ on at most $n$
vertices. Let $\varnothing\ne Q\subseteq R$, and suppose that a
nonempty set $S\subseteq V_H$ hits $Q$ in $H$, with
$D=\diam_H(S)<\infty$. Then there is a pairwise additive $+2$ spanner
$J(Q)\subseteq H$ of $(G,Q)$ satisfying
\begin{equation}\label{eq:plus2-hitting-cost}
 |E(J(Q))|=O\!\left(n+\sqrt{n|Q||S|D}\right).
\end{equation}
\end{lemma}
\begin{proof}
For each $q=(s,t)\in Q$, choose a hitting tuple $(u_q,x_q,e_q)$ with
$u_q\in S$. By \Cref{def:demand-hitting}, the edge $e_q$ is essential
to $q$, $x_q$ is an endpoint of $e_q$, and $u_qx_q\in E_H$.
Thus $x_q$ lies on every shortest path for $q$ in $H$, and
\[
 \dist_H(s,x_q)+\dist_H(x_q,t)=\dist_H(s,t)=\dist_G(s,t).
\]
It follows that
\begin{align}
 \dist_H(s,u_q)+\dist_H(u_q,t)
 &\le\dist_H(s,x_q)+1+1+\dist_H(x_q,t) \notag\\
 &=\dist_G(s,t)+2.
 \label{eq:plus2-detour}
\end{align}
Form the split demand set, as in \Cref{lem:s7-covered-deletion},
\[
 Q^*=\bigcup_{q=(s,t)\in Q}\{(u_q,s),(u_q,t)\}
       \subseteq S\times V_H,
\]
omitting repeated and trivial pairs. Hence $|Q^*|\le2|Q|$.
Choose the minimal distance preserver $H(Q^*)$ in
\Cref{eq:plus2-source} and set $J(Q)=H(Q^*)$. Concatenating its two
preserved routes through $u_q$ and using \Cref{eq:plus2-detour} gives
\[
 \dist_{J(Q)}(s,t)\le\dist_G(s,t)+2.
\]
The lower bound follows from $J(Q)\subseteq G$. The edge bound follows
from $|Q^*|\le2|Q|$. When $S$ is a singleton, its BFS-tree construction
handles all demand endpoints in the component of its root.
\end{proof}

\subsection{Hitting Sets for Long Demands}
\label{subsec:plus2-extraction}

We use \Cref{lem:technical-plus2-hitting} through the following
assigned-demand form. We retain the notation
$D_H^R(e)$ for the full essential-demand set. A chosen demand-label
$l(e)\in D_H^R(e)$ need not be the unique essential demand, so the
notation $D(e)=q$ for a private edge is not used for these labels.

\begin{corollary}
\label{lem:plus2-extraction}
Let $H=(V_H,E_H)$ be a minimal distance preserver of $(G,R)$,
retaining all $n$ vertices, and put $d_H=|E_H|/n$. Suppose that
$\dist_H(q)\ge r$ for every $q\in R$, where $6\le r\le d_H^2.$
There are nonempty sets $Q\subseteq R$ and $S\subseteq V_H$ such that
$S$ hits $Q$ in $H$ and
\begin{equation}\label{eq:plus2-extraction-guarantees}
 |Q|=\Omega(d_H^2\sqrt r),\qquad
 |S|=O(|Q|/d_H^2),\qquad \diam_H(S)\le8r.
\end{equation}
\end{corollary}
\begin{proof}
Assign each edge a demand-label $l(e)\in D_H^R(e)$, which exists by
minimality. Repeatedly remove vertices of current degree less than
$d_H/40$. Fewer than $nd_H/40$ edges are removed, so the remaining
auxiliary core $H_C=(V_C,E_C)$ satisfies
\[
 |E_C|\ge\frac{39}{40}nd_H,
 \qquad \deg_{H_C}(x)\ge\frac{d_H}{40}\quad(x\in V_C).
\]
This core need not consist of private edges. It supplies only the
selected edges and neighbors; all distances, balls, and essentiality
are measured in the unchanged graph $H$.

Apply the proof of \Cref{lem:technical-plus2-hitting} with density
parameter $d=d_H$ and length parameter $l=r$. The ball-growth proof in
\Cref{lem:growth} uses private length at least $l$ only to imply
full demand distance at least $l$. That condition holds here directly:
when $2(j-1)<r$, the endpoints of a labelled demand cannot both lie in
a radius-$(j-1)$ ball, giving the same BFS-tree covering argument.

The proofs of \Cref{lem:charge,lem:good,lem:localize} use only
essentiality to the fixed label. All edges labelled $q$ lie on a fixed
shortest path for $q$, giving the same bound of six hitting tuples per
vertex--demand incidence. Also, preserving $q$ exactly forces every
edge assigned to $q$. Thus the charging, localization, clustering, and
partial-cover arguments apply to $E_C$ with these labels. The radius
condition is $6\le r\le\min\{d_H^2,l\}$ with $l=r$. The proof gives
\Cref{eq:plus2-extraction-guarantees}, and its witness tuples show that
$S$ hits $Q$ according to \Cref{def:demand-hitting}.
\end{proof}

\begin{corollary}[Spanner for an extracted demand set]
\label{lem:plus2-repair}
Under the hypotheses of \Cref{lem:plus2-extraction}, its demand set
$Q$ has a pairwise additive $+2$ spanner $J(Q)\subseteq H$ of $(G,Q)$
with
\begin{equation}\label{eq:plus2-batch-cost}
 |E(J(Q))|=O\!\left(n+\frac{|Q|\sqrt{nr}}{d_H}\right).
\end{equation}
\end{corollary}
\begin{proof}
Apply \Cref{lem:plus2-hitting-repair} to the hitting set $S$ from
\Cref{eq:plus2-extraction-guarantees}. Its edge bound is
\[
 O\!\left(n+\sqrt{n|Q|\,(|Q|/d_H^2)\,r}\right)
 =O\!\left(n+\frac{|Q|\sqrt{nr}}{d_H}\right).
\]
\end{proof}

\subsection{Demand Removal and the Final Spanner}
\label{subsec:plus2-construction}

As in \Cref{sec:proof-main}, we keep the parameters $n,p,d,r$ fixed
while changing $R$ and its minimal distance preserver $H(R)$. Here
$d$ is the density threshold at which the construction stops. When
$H(R)$ has more than $nd$ edges, its current density
$d_R=|E(H(R))|/n$ is greater than $d$; we apply
\Cref{lem:plus2-extraction} with $d_H=d_R$.

\begin{lemma}[A parameterized demand-removal construction]
\label{prop:plus2-peeling}
For every $d\ge\sqrt6$ and $6\le r\le d^2$, there is a pairwise
additive $+2$ spanner of $(G,P)$ with edge size
\begin{equation}\label{eq:plus2-global}
 O\!\left(pr+nd+\frac{np}{d^2\sqrt r}+\frac{p\sqrt{nr}}d\right).
\end{equation}
\end{lemma}
\begin{proof}
We construct a spanning subgraph $J\subseteq G$ as follows.
\begin{enumerate}[label=\arabic*.,ref=\arabic*,leftmargin=*]
\item\label{step:plus2-short-demands}
\textbf{Preserve short demands exactly.}
Let $P_{\rm short}=\{q\in P:\dist_G(q)<r\}$. Add one $G$-shortest
path for each demand in $P_{\rm short}$ to $J$, at a cost of at most
$pr$ edges. Set $R=P\setminus P_{\rm short}$ and choose a minimal
distance preserver $H(R)\subseteq G$ of $(G,R)$. The original set $P$
and its size $p$ remain fixed.

\item\label{step:plus2-hit-batch}
\textbf{Save a spanner for a hit batch.}
While $|E(H(R))|>nd$, set $d_R=|E(H(R))|/n>d$. Every active demand
has distance at least $r$, and $r\le d^2<d_R^2$. Apply
\Cref{lem:plus2-extraction} and save the resulting subgraph $J(Q)$
from \Cref{lem:plus2-repair} by adding it to $J$.
Set $R'=R\setminus Q$, and choose a minimal distance preserver
$H(R')\subseteq H(R)$ of $(H(R),R')$. Continue with $R\gets R'$.
As in \Cref{sec:proof-main}, such an update is obtained by sequential
edge deletions that preserve the required distances; essential-demand
sets and demand-labels are recomputed in each new instance.

\item\label{step:plus2-final-preserver}
\textbf{Include the final minimal distance preserver.}
When $|E(H(R))|\le nd$, add all its edges to $J$ and stop.
\end{enumerate}

\paragraph{Correctness and termination.} For each demand $q\in R$, it always satisfies that
\begin{equation}\label{eq:plus2-invariant}
 \dist_{H(R)}(q)=\dist_G(q)\qquad(q\in R).
\end{equation}
It holds initially because $H(R)$ is a distance preserver of $(G,R)$.
The update preserves the old $H(R)$-distances exactly, so the equality
continues to hold for $R'$. Its minimality also makes $H(R')$ a
minimal distance preserver of $(G,R')$. Thus every invocation of
\Cref{lem:plus2-extraction} is in a valid current instance. Each batch
$Q$ is nonempty, so the loop terminates after at most $p$ iterations.

The demands removed in \Cref{step:plus2-short-demands} have exact
routes in $J$. Each batch removed in \Cref{step:plus2-hit-batch} has
a saved route of length at most $\dist_G(q)+2$, by
\Cref{lem:plus2-hitting-repair,eq:plus2-invariant}. Demands remaining
in \Cref{step:plus2-final-preserver} have exact routes in the final
minimal distance preserver. None of the saved subgraphs is deleted
from $J$. Therefore $J$ is a pairwise additive $+2$ spanner of $(G,P)$.
This argument does not replace parts of a demand path by several
approximate routes: each processed demand has a complete certificate
relative to its original distance in $G$.

\paragraph{Edge count.}
Let $Q_1,\ldots,Q_\tau$ be the batches removed in
\Cref{step:plus2-hit-batch}, and let $d_i$ be the corresponding value
of $d_R$. The batches are disjoint subsets of the original $P$, and
\Cref{lem:plus2-extraction} gives
\[
 \sum_{i=1}^{\tau}|Q_i|\le p,\qquad
 |Q_i|=\Omega(d_i^2\sqrt r)=\Omega(d^2\sqrt r),\qquad
 \tau=O\!\left(\frac p{d^2\sqrt r}\right).
\]
By \Cref{eq:plus2-batch-cost}, the total number of saved batch edges
is at most
\begin{align*}
 \sum_{i=1}^{\tau}O\!\left(n+\frac{|Q_i|\sqrt{nr}}{d_i}\right)
 &\le O\!\left(n\tau+\frac{\sqrt{nr}}d\sum_{i=1}^{\tau}|Q_i|\right)\\
 &=O\!\left(\frac{np}{d^2\sqrt r}+\frac{p\sqrt{nr}}d\right).
\end{align*}
Adding the at most $pr$ initial path edges and the at most $nd$ final
edges proves \Cref{eq:plus2-global}. The same bound holds if no batch
is extracted.
\end{proof}
\paragraph{Complete the Proof of \Cref{thm: main +2 spanner}}

For large demand sets, we use the following classical all-pairs additive spanner bound.

\begin{theorem}[Elkin--Peleg~{\cite[Theorem~3.22(1)]{EP04}}]
\label{lem:plus2-all-pairs}
Every $n$-vertex undirected unweighted graph has an all-pairs additive
$+2$ spanner with $O(n^{3/2})$ edges.
\end{theorem}

\begin{proof}[Proof of \Cref{thm: main +2 spanner}]
Put $T=n^{9/10}p^{2/5}$. If $p<n^{2/3}$, \Cref{lm: CE06} gives a
minimal distance preserver of size $O(n+\sqrt n\,p)$, and
\[
 \frac{\sqrt n\,p}{T}=n^{-2/5}p^{3/5}\le1.
\]
If $p>n^{3/2}$, by \Cref{lem:plus2-all-pairs}, we have
$n^{3/2}\leq T$. It remains to consider
$n^{2/3}\le p\le n^{3/2}$. Fix the parameters
\begin{equation}\label{eq:plus2-parameters}
 d=3n^{-1/10}p^{2/5},\qquad r=\max\{6,\sqrt n/d\}.
\end{equation}
Here $d\ge3n^{1/6}$, so $d^2\ge6$ and $d^3\ge27\sqrt n$.
Consequently $6\le r\le d^2$, allowing
\Cref{prop:plus2-peeling}. The maximum with six ensures the technical
radius condition even when the balanced value $\sqrt n/d$ is small.

We bound the four costs in \Cref{eq:plus2-global} using the actual
radius in \Cref{eq:plus2-parameters}. First, $nd=3T$. Since
$r\ge\sqrt n/d$,
\begin{equation}\label{eq:plus2-balanced-cost}
 \frac{np}{d^2\sqrt r}\le\frac{pn^{3/4}}{d^{3/2}}=3^{-3/2}T.
\end{equation}
Also $\sqrt r\le\sqrt6+n^{1/4}d^{-1/2}$, hence
\[
 \frac{p\sqrt{nr}}d
 \le\frac{\sqrt6\,p\sqrt n}d+\frac{pn^{3/4}}{d^{3/2}}.
\]
The second term is bounded in \Cref{eq:plus2-balanced-cost}. For the
first, $p\le n^{3/2}$ gives
\[
 \frac{p\sqrt n}d=\frac13n^{3/5}p^{3/5}\le T/3.
\]
Finally,
\[
 pr\le6p+\frac{p\sqrt n}d=O(T),
\]
since $p/T=n^{-9/10}p^{3/5}\le1$ in this range. Substitution into
\Cref{eq:plus2-global} proves the claimed bound for all demand counts.
\end{proof}

\section{More Results on Additive Spanner Bounds}
\label{sec: +4}

We derive the remaining bounds in \Cref{tab:additive-spanners} from
\Cref{thm: main} and classical preserver bounds using the path-sampling and clustering framework
of~\cite{Kav17}, with the constant-fraction completion argument
of~\cite{Bod22}. We give the reductions explicitly.
Throughout, $G=(V,E)$ is an undirected unweighted graph with $n$ vertices,
$P$ is a demand set of size $p$, and $J\subseteq G$ denotes an additive
spanner, as in \Cref{def:pairwise-additive-spanner}.
All subgraphs retain $V$, and all distance guarantees are relative to $G$.
We may discard trivial and unreachable demands and assume $n\ge2$;
empty demand or source sets require no edges.

\begin{definition}[Source-sensitive and sourcewise spanners]
\label{def:s8-sourcewise}
Let $S\subseteq V$ have size $\sigma$ and let $\beta\ge0$ be an integer.
For $P\subseteq S\times V$,
a source-sensitive additive $+\beta$ spanner is a subgraph $J\subseteq G$
such that
\[
 \dist_J(s,t)\le \dist_G(s,t)+\beta\qquad((s,t)\in P).
\]
A complete sourcewise additive $+\beta$ spanner satisfies this inequality
for every reachable pair in $S\times V$.
For unordered demands, $P\subseteq S\times V$ means that each demand
has been oriented with its first endpoint in $S$.
\end{definition}

We use the following exact-preserver bounds:
\begin{align}
 &O(n+n^{8/9}q^{4/9}) &&\text{for }q\le n,
       \label{eq:s8-new-exact}\\
 &O(n+\sqrt n\,q) &&\text{for all }q.
       \label{eq:s8-CE-exact}
\end{align}
Here \eqref{eq:s8-CE-exact} is \Cref{lm: CE06}, and
\eqref{eq:s8-new-exact} follows from \Cref{thm: main}, using
\eqref{eq:s8-CE-exact} when $q<\sqrt n$.
In particular, at most $\sqrt n$ demands have an $O(n)$-edge exact preserver.

\subsection{Two common reductions}

\begin{lemma}[Sampling reduction]\label{lem:s8-sampling}
Let $p>0$, $d\ge1$, and let $\beta\ge0$ be an integer.
Each of the following constructions provides a subgraph $J_0\subseteq G$
with $O(nd)$ edges and a set $R\subseteq V$ with $|R|\le p/d^2$.
\begin{enumerate}[label=(\roman*)]
\item If $P\subseteq S\times V$, adjoining any additive $+\beta$ spanner
for the reachable pairs in $S\times R$ satisfies at least half of $P$
with additive error $\beta+2$.
\item For arbitrary $P$, adjoining any additive $+\beta$ spanner for
the reachable pairs in $R\times R$ satisfies at least half of $P$
with additive error $\beta+4$.
\end{enumerate}
The choices of $J_0$ and $R$ may differ between the two constructions.
\end{lemma}

\begin{proof}
Put $\Delta=\lceil4d\rceil$ and $\ell=\lceil n\Delta/p\rceil$.
Form $B\subseteq G$ by retaining $\Delta$ arbitrary incident edges at
each vertex, or all its incident edges if fewer exist.
Thus $|E(B)|\le n\Delta$.
For each demand $(s,t)\in P$, choose an arbitrary shortest $s$--$t$
path in $G$, denote it by $\pi(s,t)$, and traverse it from $s$ to $t$.
In (i), use the orientation with $s\in S$; in (ii), either orientation
may be chosen. These paths remain fixed throughout this construction.
The argument uses only shortestness of each individual path, so the
choices for different demands may be made independently.
Throughout this proof, a missing edge means an edge absent from this fixed $B$.

For (i), purchase all missing edges of a path if there are fewer than $\ell$;
otherwise purchase its last $\ell-1$ missing edges.
For (ii), purchase all missing edges if there are fewer than $2\ell-1$;
otherwise purchase its first and last $\ell-1$ missing edges.
Here the first and last missing edges are ordered along $\pi(s,t)$
from $s$ to $t$. All purchases are made relative to $B$, even if an
edge has already been purchased for another demand.
Let $J_0$ be $B$ together with these purchases.
In either case their total cost is at most
$n\Delta+2p(\ell-1)<3n\Delta=O(nd)$.

For a path not bought entirely, in (i) consider the heads of its last
$\ell$ missing edges; in (ii) also consider the tails of its first
$\ell$ missing edges. Every endpoint of a missing edge has at least
$\Delta$ neighbors in $B$: a vertex of degree at most $\Delta$ would
have retained all its incident edges.
For a demand $q$, let $A_q^+$ be the union of the $B$-neighbors of its
chosen heads, and in (ii) let $A_q^-$ be the analogous union for its
chosen tails. Each list has $\ell$ distinct vertices, since the chosen
path is simple. Any vertex is adjacent to at most three vertices of a
shortest path: the subpath between its first and last neighbors has
length at most two, using the route through that vertex.
Counting incidences between the list and its $B$-neighbors therefore gives
\[
 |A_q^+|\ge\ell\Delta/3,
 \qquad |A_q^-|\ge\ell\Delta/3\quad\text{in (ii)}.
\]
The suffix from each chosen head to $t$, and in (ii) the prefix from
$s$ to each chosen tail, belongs to $J_0$.
In (ii) every chosen tail precedes every chosen head, even when the
two lists of missing edges share their middle edge.

If every path was bought entirely, take $R=\emptyset$; then $J_0$
already preserves every demand exactly.
Otherwise one of these neighborhoods exists, and its size is at most
$n$, so $\ell\Delta\le3n$. Sample
$m=\lceil12n/(\ell\Delta)\rceil$ vertices independently and uniformly
with replacement, and let $R$ be their set. Then
\[
 |R|\le m\le\frac{15n}{\ell\Delta}
       \le\frac{15p}{\Delta^2}\le\frac{p}{d^2}.
\]
For each indicated neighborhood $A$, the probability that it contains
no sampled vertex is
\[
 \Pr(R\cap A=\emptyset)
   =\left(1-\frac{|A|}{n}\right)^m
   \le\exp\!\left(-\frac{m|A|}{n}\right)\le e^{-4}.
\]
In (i), a sampled neighbor $z$ of a chosen head $v$ gives, after adding
the auxiliary spanner, a route of length at most
\[
 \dist_G(s,z)+\beta+1+\dist_G(v,t)
 \le\dist_G(s,t)+\beta+2.
\]
Here $zv\in E(B)$ gives $\dist_G(s,z)\le\dist_G(s,v)+1$,
and the chosen path gives
$\dist_G(s,v)+\dist_G(v,t)=\dist_G(s,t)$.
In (ii), sampled neighbors $a,b$ of a chosen tail $u$ and head $v$
give a route of length at most
\[
 \dist_G(s,u)+1+\dist_G(a,b)+\beta+1+\dist_G(v,t)
 \le\dist_G(s,t)+\beta+4.
\]
The bound on the route length uses the auxiliary distance guarantee.
The displayed inequality follows from
$\dist_G(a,b)\le1+\dist_G(u,v)+1$ and the fact that $u$ precedes
$v$ on a shortest $s$--$t$ path.
In (i) it suffices that $R$ meets $A_q^+$; in (ii) it suffices that
$R$ meets both $A_q^-$ and $A_q^+$. A union bound shows that each
demand not bought entirely has these witnesses with probability at
least $1-2e^{-4}>1/2$. Demands bought entirely always succeed.
Let $X$ count demands with such certificates. By linearity of expectation,
\[
                    \mathbb E[X]\ge(1-2e^{-4})p>p/2.
\]
Hence some sampling outcome certifies at least half the demands.
The certificates use only the stated auxiliary distance guarantees,
so they work for any auxiliary spanner satisfying those guarantees.
The edge and sample-size bounds hold for every outcome, proving the lemma.
\end{proof}

\begin{lemma}[Completion from a constant fraction]\label{lem:s8-completion}
Fix an additive error $\beta$ and a source set $S$, if applicable.
Suppose that, whenever $p>\sqrt n$, at least half the demands can be
satisfied with $O(f(n,\sigma,p))$ edges, where
$f(n,\sigma,p)=A(n,\sigma)p^\eta$, with fixed $\eta>0$, and
$f(n,\sigma,p)\ge n$ for these demand counts.
If the construction's hypotheses persist on demand subsets, then all
demands have an additive $+\beta$ spanner with $O(n+f(n,\sigma,p))$ edges.
\end{lemma}

\begin{proof}
If $p\le\sqrt n$, \eqref{eq:s8-CE-exact} gives an exact preserver
with $O(n)$ edges. Otherwise repeat the constant-fraction construction
on the unsatisfied demands in the original graph $G$, keeping $S$ fixed.
After $i$ rounds, at most $p/2^i$ demands remain. Since $\eta>0$, the
successive costs sum geometrically:
\[
 \sum_{i\ge0}O\!\left(A(n,\sigma)(p/2^i)^\eta\right)
 =O\!\left(\frac{f(n,\sigma,p)}{1-2^{-\eta}}\right)
 =O(f(n,\sigma,p)).
\]
Finish with \eqref{eq:s8-CE-exact} when at most $\sqrt n$ demands remain.
Every round supplies complete routes for its satisfied demands relative
to their distances in $G$. Their union retains all these routes, so the
additive error remains $\beta$ throughout the completion.
\end{proof}

\subsection{Pairwise and source-sensitive bounds}

\begin{theorem}\label{thm: main +4 spanner}
For every set $P$ of $p\le n^{7/6}$ demand pairs, $(G,P)$ admits a
pairwise additive $+4$ spanner with
$O(n+n^{24/25}p^{8/25})$ edges.
\end{theorem}

\begin{proof}
The case $p\le\sqrt n$ follows from \eqref{eq:s8-CE-exact}.
Otherwise set $d=n^{-1/25}p^{8/25}>1$ and apply
\Cref{lem:s8-sampling}(ii) with an exact auxiliary preserver.
Its demand count is
\[
 q\le |R|^2\le p^2/d^4=n^{4/25}p^{18/25}\le n.
\]
Hence \eqref{eq:s8-new-exact} applies and costs
$O(n+n^{8/9}(p^2/d^4)^{4/9})=O(nd)$.
The sampling step therefore satisfies half the demands at cost
$O(nd)=O(n^{24/25}p^{8/25})$.
Here $nd>n$, and the range $p\le n^{7/6}$ persists when demands are
removed. Thus \Cref{lem:s8-completion}, with $\eta=8/25$, proves the claim.
\end{proof}

\begin{theorem}[Source-sensitive bounds]
\label{thm: source-sensitive additive}
\label{thm: main +2 spanner source-sensitive}
Let $P\subseteq S\times V$ and $|S|=\sigma$.
For every fixed integer $k\ge1$, if $p\sigma^k\le n^{1+2k/3}$, then
$(G,P)$ admits an additive $+2k$ spanner with
\[
 O_k(n+F_k(n,\sigma,p))\quad\text{edges},\qquad
 F_k(n,\sigma,p)=
 n^{(8+8k)/(9+8k)}(p\sigma^k)^{4/(9+8k)}.
\]
In particular, for $k=1,2,3$ the nonlinear terms are
$n^{16/17}(p\sigma)^{4/17}$,
$n^{24/25}(p\sigma^2)^{4/25}$, and
$n^{32/33}(p\sigma^3)^{4/33}$, respectively, with the ranges in
\Cref{tab:additive-spanners}. No additional restriction $p\le n$ is needed.
\end{theorem}

\begin{proof}
Extend the definition of $F_k$ to $k=0$; then
\eqref{eq:s8-new-exact} is the induction base, valid for $p\le n$.
Assume $k\ge1$. The case $p\le\sqrt n$ follows from
\eqref{eq:s8-CE-exact}. Otherwise put
\[
 a=9+8k,\qquad M=p\sigma^k,\qquad d=n^{-1/a}M^{4/a}>1.
\]
Apply \Cref{lem:s8-sampling}(i) and construct a $+2(k-1)$ spanner
for all reachable pairs in $S\times R$.
If their number is $q$, then
\[
 q\sigma^{k-1}\le M/d^2
 =n^{2/a}M^{(a-8)/a}\le n^{1+2(k-1)/3}.
\]
Indeed, substituting $M\le n^{1+2k/3}$ into the middle expression
gives $n^{(2k+1)/3}=n^{1+2(k-1)/3}$.
Thus the induction hypothesis applies to the auxiliary demands,
including $q\le n$ when $k=1$. If no auxiliary demand is reachable,
the empty auxiliary subgraph suffices.
Since $d^a=M^4/n$, its nonlinear cost satisfies
\[
 F_{k-1}(n,\sigma,q)
 \le n^{8k/(8k+1)}(M/d^2)^{4/(8k+1)}=nd.
\]
The auxiliary linear term $O_k(n)$ is absorbed by $O_k(nd)$, since
$d>1$. Also $nd=F_k(n,\sigma,p)$. The sampling step adds two to the
error, so at least half the demands have additive error $2k$ at this cost.
For fixed $n,\sigma,k$, the function $F_k$ is proportional to
$p^{4/(9+8k)}$, and the range $p\sigma^k\le n^{1+2k/3}$ persists on
demand subsets. Therefore \Cref{lem:s8-completion} completes the induction.
Only the level-zero auxiliary calls require at most $n$ pairs; the
final exact completion uses at most $\sqrt n$ pairs.
\end{proof}

\begin{lemma}[Classical-input source-sensitive bounds]
\label{lem:s8-classical-source}
For every fixed integer $k\ge0$ and $P\subseteq S\times V$, there is an
additive $+2k$ spanner with
\[
 O_k(n+C_k(n,\sigma,p))\quad\text{edges},\qquad
 C_k(n,\sigma,p)=n^{1-1/(4k+2)}(p\sigma^k)^{1/(2k+1)}.
\]
There is also such a spanner with
$O_k(n+n(p\sigma^k)^{1/(2k+3)})$ edges when
$p\sigma^k\le n^{1+2k/3}$.
\end{lemma}

\begin{proof}
For the first bound, induct on $k$. The base $k=0$ is
$C_0(n,\sigma,p)=\sqrt n\,p$, supplied by \eqref{eq:s8-CE-exact}
without a restriction on $p$. Suppose $k\ge1$ and $p>\sqrt n$,
write $M=p\sigma^k$, and choose
$d=n^{-1/(4k+2)}M^{1/(2k+1)}>1$.
Apply \Cref{lem:s8-sampling}(i). If $q$ is the number of reachable
auxiliary pairs in $S\times R$, then $q\sigma^{k-1}\le M/d^2$.
The induction hypothesis and $d^{2k+1}=M/\sqrt n$ give
\[
 C_{k-1}(n,\sigma,q)
 \le n^{1-1/(4k-2)}(M/d^2)^{1/(2k-1)}=nd.
\]
Consequently, an additive $+2(k-1)$ auxiliary spanner and the sampling
subgraph together satisfy half the demands with error $2k$ and cost
$O_k(nd)$. Since $nd=C_k(n,\sigma,p)$ is a positive power of $p$,
\Cref{lem:s8-completion} proves the first bound. There is no restriction
on the exact auxiliary demand count.

For the second bound, the base $k=0$ is the exact-preserver theorem of~\cite{BVW21}, which
gives $O(n^{2/3}q^{2/3}+nq^{1/3})=O(nq^{1/3})$ for $1\le q\le n$.
For $k\ge1$ and $p>\sqrt n$, choose $d=M^{1/(2k+3)}>1$ and again
apply \Cref{lem:s8-sampling}(i). The auxiliary product satisfies
\[
 q\sigma^{k-1}\le M/d^2=M^{(2k+1)/(2k+3)}
                    \le n^{(2k+1)/3}=n^{1+2(k-1)/3},
\]
so the induction hypothesis applies, including the base's $q\le n$
condition when $k=1$. Its nonlinear cost is at most
$n(M/d^2)^{1/(2k+1)}=nd$. The linear term is absorbed since $d>1$.
The sampling step therefore satisfies half the demands with error
$2k$ at cost $O_k(nd)$. The range persists on demand subsets and
$nd=n(p\sigma^k)^{1/(2k+3)}$ is a positive power of $p$, allowing
\Cref{lem:s8-completion}. In both inductions, the cases
$p\le\sqrt n$ are handled directly by \eqref{eq:s8-CE-exact}.
\end{proof}

\subsection{Complete sourcewise bounds}

\begin{lemma}[Clustering reduction]\label{lem:s8-clustering}
For every integer $\Delta\ge1$ there are $J_0\subseteq G$ and
$R\subseteq V$ with $|E(J_0)|=O(n\Delta)$ and $|R|\le n/\Delta$
such that the following holds for every source set $S$.
Adjoining any additive $+\beta$ spanner for the reachable pairs in
$S\times R$ produces a complete sourcewise additive $+(\beta+2)$ spanner.
\end{lemma}

\begin{proof}
Initially all vertices are unclustered. While a vertex has at least
$\Delta$ unclustered neighbors, select exactly $\Delta$ of them as a
cluster, assign this vertex as their center, and mark these neighbors
clustered. A center need not itself be unclustered and may be reused.
Let $R$ be the set of centers, so $|R|\le n/\Delta$.
Keep all cluster-star edges, at most $n$ in total, and every edge
incident to the final unclustered set $U$.
At termination $|N_G(v)\cap U|<\Delta$ for every $v$, so these latter
edges number at most $\sum_v|N_G(v)\cap U|<n\Delta$.
Let $J_0$ be their union.

For a reachable $(s,t)\in S\times V$, choose any shortest $s$--$t$
path $\pi(s,t)$ in $G$ and traverse it from $s$ to $t$.
If it is contained in $J_0$, it is already preserved.
Otherwise let $v$ be the head of its last missing edge.
Both endpoints of that edge are clustered, since every edge incident
to an unclustered vertex was retained. In particular, $v$ is clustered,
and the subpath from $v$ to $t$ lies in $J_0$ by the choice of the last
missing edge.
If $z\in R$ is its center, the auxiliary $s$--$z$ route, the edge $zv$,
and this suffix have total length at most
\[
 \dist_G(s,z)+\beta+1+\dist_G(v,t)
 \le\dist_G(s,t)+\beta+2.
\]
Here $\dist_G(s,z)\le\dist_G(s,v)+1$, because $zv\in E$,
and the chosen path gives
$\dist_G(s,v)+\dist_G(v,t)=\dist_G(s,t)$.
The path can be chosen independently for each pair, and the same
$J_0$ and $R$ work for every source set $S$.
\end{proof}

\begin{theorem}[Complete sourcewise bounds]
\label{thm: complete-sourcewise additive}
For every fixed integer $k\ge2$ and source set $S$ of size
$\sigma\le n^{(2k-1)/(3k)}$, there is a complete sourcewise additive
$+2k$ spanner with
\[
 O_k\!\left(n+n^{(8+8k)/(5+8k)}\sigma^{4k/(5+8k)}\right)
\]
edges. For $k=2$ this gives $O(n+n^{8/7}\sigma^{8/21})$ edges when
$\sigma\le n^{1/2}$. For $k=3$ it gives
$O(n+n^{32/29}\sigma^{12/29})$ edges when $\sigma\le n^{5/9}$.
\end{theorem}

\begin{proof}
Set $d=n^{3/(8k+5)}\sigma^{4k/(8k+5)}$ and
$\Delta=\lceil d\rceil$. Use \Cref{lem:s8-clustering} and apply
\Cref{thm: source-sensitive additive} at level $k-1$ to the reachable
pairs in $S\times R$. Their count $q\le n\sigma/\Delta$ satisfies
\[
 q\sigma^{k-1}\le n\sigma^k/d\le n^{1+2(k-1)/3},
\]
where the last inequality follows from the assumed bound on $\sigma$.
The auxiliary nonlinear cost is at most
\[
 F_{k-1}(n,\sigma,q)
 \le n^{8k/(8k+1)}(n\sigma^k/d)^{4/(8k+1)}=nd.
\]
The equality in the last display follows from
$d^{8k+5}=n^3\sigma^{4k}$. The auxiliary linear term is $O_k(n)$.
The clustering cost is $O(n\Delta)=O(n+nd)$, and its attachment adds
two to the auxiliary error $2(k-1)$. Thus the final error is $2k$;
substituting $d$ into $nd$ proves the stated edge bound.
\end{proof}

\begin{corollary}[A classical-input sourcewise bound]
\label{cor: CE-sourcewise additive}
For every source set $S$ of size $\sigma$, there is a complete sourcewise
additive $+6$ spanner with $O(n+n^{13/12}\sigma^{1/2})$ edges.
\end{corollary}

\begin{proof}
Take $\Delta=\lceil n^{1/12}\sigma^{1/2}\rceil$ in
\Cref{lem:s8-clustering}. Apply the first bound of
\Cref{lem:s8-classical-source} with $k=2$ to $S\times R$.
Since $q\le n\sigma/\Delta$, its nonlinear cost is
\[
 C_2(n,\sigma,q)=n^{9/10}(q\sigma^2)^{1/5}
 \le n^{11/10}\sigma^{3/5}\Delta^{-1/5}
 \le n^{13/12}\sigma^{1/2}.
\]
The clustering cost has the same bound, up to $O(n)$, and the error is
$4+2=6$.
\end{proof}

Comparing the two complete-sourcewise $+6$ bounds, we have
\[
 \frac{n^{32/29}\sigma^{12/29}}{n^{13/12}\sigma^{1/2}}
 =n^{7/348}\sigma^{-5/58}.
\]
Thus, for $\sigma=n^a$ with fixed $7/30<a<14/31$, the $k=3$ bound
of \Cref{thm: complete-sourcewise additive} gives a strict polynomial
improvement over both \Cref{cor: CE-sourcewise additive} and the previous
bounds listed in \Cref{tab:additive-spanners}.

\paragraph{Classical comparison terms in \Cref{tab:additive-spanners}.}
The source-sensitive CE and BW terms follow by setting $k=1,2,3$ in
\Cref{lem:s8-classical-source}.
The pairwise $+4$ term $O(n+n^{9/10}p^{2/5})$ also follows from
\Cref{lem:s8-sampling}(ii) using \eqref{eq:s8-CE-exact} and
$d=n^{-1/10}p^{2/5}$, followed by \Cref{lem:s8-completion}.
The complete-sourcewise $+6$ term $O(n+n^{15/14}\sigma^{4/7})$
follows directly from the first bound of \Cref{lem:s8-classical-source}
with $k=3$ and $p\le n\sigma$, since
\[
 C_3(n,\sigma,p)
 \le n^{13/14}(n\sigma\cdot\sigma^3)^{1/7}
 =n^{15/14}\sigma^{4/7}.
\]
This bound applies to every source set and uses
\eqref{eq:s8-CE-exact} through three levels of the sampling reduction,
with constant-fraction completion at each level.
For the remaining two derived partial-demand comparisons, use
\Cref{lem:s8-sampling}(i).
The subset $+2$ theorem~\cite{Bod22} supplies a spanner for $T\times T$
with $O(n\sqrt{|T|})$ edges. Applying it to $T=S\cup R$ costs
\[
 O\!\left(n\sqrt{\sigma+p/d^2}\right).
\]
With $d=\max\{\sigma^{1/2},p^{1/4}\}$ this is $O(nd)$.
Keep $d$ fixed and repeat on the unsatisfied demands until at most
$\sqrt n$ remain, then finish with \eqref{eq:s8-CE-exact}.
Since the number of demand pairs is at most $n^2$, there are
$O(\log n)$ rounds. This gives the $+4$ comparison bound with
$\widetilde O(nd)$ edges.

For the $+6$ comparison, use the rectangular $+4$ theorem of~\cite{Kav17}.
Its auxiliary cost is $O(n(\sigma|R|)^{1/4})$.
Taking $d=(\sigma p)^{1/6}$ bounds this cost by $O(nd)$;
\Cref{lem:s8-completion} gives $O(n+n(\sigma p)^{1/6})$ edges.
In these applications $p\le\sqrt n$ is handled directly by
\eqref{eq:s8-CE-exact}.

\paragraph{Acknowledgments.}
We thank Greg Bodwin for helpful discussions at the early stage of the project. We also thank the anonymous reviewers for their helpful feedback and for pointing out potential new research directions.

\paragraph{AI methodology statement.}
We used ChatGPT 6-Astra to assist in developing the results in
\Cref{sec:plus2-pairwise,sec: +4}. 
All work concerning \Cref{thm: main}, including its development and proof, was carried out entirely by the authors without AI assistance.
The authors wrote the entire paper and take full responsibility for all of its contents, including the correctness of the results and proofs.

\bibliographystyle{alpha}
\bibliography{ref}
\appendix

\section{Remaining Proofs}\label{appendix:remaining proofs}

\begin{proof}[Proof of \Cref{lem: generalized p sigma d}]
    We adapt the distance-difference argument of \cite{BVW21}, using essential edges to handle an arbitrary minimal distance preserver.
    For every demand $q=(s,t)\in P$, fix a shortest path $\pi(q)$ in $H$, directed from $s$ to $t$.
    Since $H$ preserves the demand distances, each $\pi(q)$ is also a shortest path in $G$.
    By minimality, choose a demand-label $l(e)$ for every edge $e\in E_H$, and orient $e$ as it appears on $\pi(l(e))$.
    We call the source of $l(e)$ the source of the oriented edge $e$.

    If an edge $e=(a,x)$ has source $u$, then
    $\dist_H(u,x)=\dist_G(u,x)$, and every shortest $u$--$x$ path in $H$ ends with $e$.
    Indeed, the prefix of $\pi(l(e))$ ending at $x$ proves the distance equality.
    Any shortest $u$--$x$ path avoiding $e$ could replace this prefix, contradicting that $e$ is essential to $l(e)$.
    Consequently, at most one edge with a given source enters any vertex.

    Fix a demand $q=(s,t)\in P$ and a source $u\in S$.
    Let $F(q,u)$ consist of the edges with source $u$ entering a vertex of $\pi(q)$, other than the edge preceding that vertex on $\pi(q)$.
    We claim that
    \[
        |F(q,u)|\leq 2\lfloor d\rfloor+1.
    \]
    For each head $x$ of an edge in $F(q,u)$, define
    \[
        \Delta(x)=\dist_G(u,x)-\dist_G(s,x).
    \]
    The triangle inequality in $G$ gives $-d\leq\Delta(x)\leq d$, so there are at most $2\lfloor d\rfloor+1$ possible integer values.
    Suppose two distinct heads $x,y$ have the same value, with $x$ preceding $y$ on $\pi(q)$.
    Writing $|\pi(q)[x,y]|$ for the length of the subpath from $x$ to $y$, we have
    \[
        \dist_G(u,x)+|\pi(q)[x,y]|=\dist_G(u,y).
    \]
    A shortest $u$--$x$ path in $H$ followed by $\pi(q)[x,y]$ is therefore a shortest $u$--$y$ path in $H$.
    It avoids the edge $f\in F(q,u)$ entering $y$: the first part has length strictly less than $\dist_G(u,y)$ and hence cannot visit $y$, while the second part does not contain $f$, since $f$ is not the predecessor edge at $y$ on $\pi(q)$.
    This contradicts the preceding property of $f$ and proves the claim.

    Let $\deg_H^{\mathrm{in}}(x)$ be the indegree in this orientation, and set
    \[
        b=\sum_{x\in V_H}\binom{\deg_H^{\mathrm{in}}(x)}{2}.
    \]
    Fix a total ordering of $E_H$.
    For each pair of distinct edges $e,f$ entering the same vertex, with $e$ preceding $f$ in this ordering, charge the pair to $(l(e),u,f)$, where $u$ is the source of $f$.
    The edge $e$ is the predecessor edge on $\pi(l(e))$ at their common head, so $f\in F(l(e),u)$.
    This charge is injective, since $l(e)$ and the head of $f$ uniquely determine $e$.
    Thus
    \[
        b\leq\sum_{q\in P}\sum_{u\in S}|F(q,u)|
        \leq p\sigma(2\lfloor d\rfloor+1).
    \]
    If $\sigma\geq2$, then $d\geq1$ and hence $b=O(p\sigma d)$.
    If $\sigma=1$, every vertex has indegree at most one, so $b=0$; in particular, the same bound holds when $d=0$.

    Finally, restrict the orientation to $H[U]$, and write $m=|E_H^U|$ and $a_x=\deg_{H[U]}^{\mathrm{in}}(x)$.
    The case $k=0$ is immediate. Otherwise, by Cauchy--Schwarz,
    \[
        m^2=\left(\sum_{x\in U}a_x\right)^2
        \leq k\sum_{x\in U}a_x^2
        =k\left(m+2\sum_{x\in U}\binom{a_x}{2}\right)
        \leq k(m+2b).
    \]
    It follows that
    \[
        |E_H^U|=m\leq k+\sqrt{2kb}
        =O\!\left(k+\sqrt{kp\sigma d}\right),
    \]
    as desired.
\end{proof}

\end{document}